 \documentstyle{article}

  \font\smallbf=cmbx7
  \font\bfcal=cmbsy10

 \def\sgn{\mathop{\rm sgn}\nolimits}

\catcode`\@=11

   \renewcommand{\section}%
   {\setcounter{equation}{0}\@startsection {section}{1}{\z@}{-3.5ex plus -1ex
   minus -.2ex}{2.3ex plus .2ex}{\Large\bf}}

   \newcommand{\beq}{\begin{equation}}
   \newcommand{\eeq}{\end{equation}}
   \newcommand{\beqs}{\arraycolsep1.5pt\begin{eqnarray}}
   \newcommand{\eeqs}{\arraycolsep5pt\end{eqnarray}}
   \newcommand{\beqsn}{\arraycolsep1.5pt\begin{eqnarray*}}
   \newcommand{\eeqsn}{\end{eqnarray*}\arraycolsep5pt}
   \newcommand{\bmatrix}{\arraycolsep5pt\begin{array}}
   \newcommand{\ematrix}{\arraycolsep1.5pt\end{array}}

\catcode`\@=12

\newtheorem{prop}{Proposition}[section]
\newtheorem{lemma}{Lemma}[section]

\begin{document}



\title{PROPERTIES OF SOLUTIONS OF THE KPI EQUATION
\thanks{Work supported in part by Ministero delle Universit\`a e
della Ricerca Scientifica e Tecnologica, Italia}}
\author{M.~Boiti \and
F.~Pempinelli \and A.~Pogrebkov\thanks{Permanent address:  Steklov
Mathematical Institute, Vavilov Str. 42, Moscow 117966, GSP-1, RUSSIA;
e-mail POGREB@QFT.MIAN.SU}\\ Dipartimento di Fisica dell'Universit\`a and
Sezione INFN\thanks{e-mail: BOITI@LECCE.INFN.IT and
PEMPI@LECCE.INFN.IT}\\ 73100 Lecce, ITALIA}
\date{November 3, 1993}
\maketitle

\begin{abstract}
The Kadomtsev--Petviashvili I (KPI) is considered as a useful laboratory for
experimenting new theoretical tools able to handle the specific features of
integrable  models in $2+1$ dimensions.
The linearized version of the  KPI equation is
first considered by solving the initial value problem for different classes
of initial data. Properties of the solutions in different cases are analyzed
in details. The obtained results are used as a guideline for studying the
properties of the solution $u(t,x,y)$ of the Kadomtsev--Petviashvili I (KPI)
equation with given initial data $u(0,x,y)$ belonging to the Schwartz space.
The spectral theory associated to KPI is studied in the space of the Fourier
transform of the solutions. The variables $p=\{p_1,p_2\}$ of the Fourier
space are shown to be the most convenient spectral variables to use for
spectral data. Spectral data are shown to decay rapidly at large $p$ but to
be discontinuous at $p=0$. Direct and inverse problems are solved with
special attention to the behaviour of all the quantities involved in the
neighborhood of $t=0$ and $p=0$. It is shown in particular that the solution
$u(t,x,y)$ has a time derivative discontinuous at $t=0$ and that at any
$t\not=0$ it does not belong to the Schwartz space no matter how small in
norm and rapidly decaying at large distances the initial data are chosen.
\end{abstract}


\section{Introduction}

We consider the initial value problem for the Kadomtsev--Petviashvili
equation~\cite{kp,brown,ablseg} in its version called KPI
\beqs
(u_{t}-6uu_{x}+u_{xxx})_{x}&=&3u_{yy}\,,\qquad u=u(t,x,y)\,,\nonumber\\
u(0,x,y)=u(x,y)\,,
\label{kp1}
\eeqs
for $u(t,x,y)$ real.
Already in 1974~\cite{dryuma} it has been acknowledged to be integrable
since it can be associated to a linear spectral problem and, precisely, to
the non--stationary Schr\"odinger equation
\beq
\label{Schrodin}
(-i\partial _{y}+\partial ^{2}_{x}-u(x,y))\Phi =0.
\eeq
However, to building a complete and coherent theory for
the spectral transform of the potential $u(x,y)$ in (\ref{Schrodin}) that
could be used to linearize the initial value problem of (\ref {kp1})
resulted to be unexpectedly difficult. The real breakthrough has been the
discover that the problem is solvable via a non local Riemann--Hilbert
formulation~\cite{ZManakov,FokasAblowitz}.

Successively other progresses have been made. The characterization
problem for the spectral data was solved in~\cite{KP}.  The extension of
the spectral transform to the case of a potential $u(x,y)$ approaching to
zero in every direction except a finite number has been faced
in~\cite{total}. The questions of the associated conditions (often called
`constraints') and how to choose properly $\partial^{-1}_x$ in the evolution
form of KPI
\beq
\label{kp1evol}
u_{t}(t,x,y)-6u(t,x,y)u_{x}(t,x,y)+u_{xxx}(t,x,y)=3\partial^{-1}_x\,
u_{yy}(t,x,y)
\eeq
have been studied in~\cite{ablvill}. Some relevant points on
constraints were known earlier, but have not been published~\cite{Manakov}.
See also~\cite{Bakurov}. In the
article~\cite{Zhou} some rigorous results on the existence and uniqueness of
solutions of the direct and inverse problem have been proved.

In this article, as well as in the previous one~\cite{kpshort} by the same
authors, the description
of the KPI equation by means of the inverse spectral transform is revisited.
We think that this is necessary since some specific properties of the
solutions
of the initial value problem~(\ref {kp1}) and, more generally, of the
$2+1$--integrable equations
are not properly (or in any case not exhaustively) described in the
literature. Moreover, these properties are specially interesting since they
seem to be peculiar of the integrable
equations in $2+1$ dimensions. Our study is essentially based on some
new
theoretical tools in the theory of the spectral transform developed
in~\cite{total,resreg,nsreshi}.

In this article we consider the equation~(\ref {kp1}) for initial data
$u(x,y)$
belonging to the Schwartz space ${\cal S}$. The main results we get as
regards the properties of the solutions and the theory of the spectral
transform can be summarized in the following points (the integrations when
it is not differently indicated are performed all along the real axis from
$-\infty$ to $+\infty$).
\begin{enumerate}
\item
The solution $u(t,x,y)$ immediately leaves the Schwartz space since at any
time $t\not=0$ develops a tail slowly decreasing for  $x\to t\infty$
according to the following formula
\beq
\label{asympt1}
u(t,x,y)= {-1\over 4\pi \sqrt{3tx}|x|}\!\int\!\!\!\int\!\!dx'dy'\,u(x',y')+
o\bigl(|x|^{-3/2}\bigr)\,.
\eeq
Only in the opposite direction, i.e. for $x\to -t\infty$, it decreases
rapidly.
\item
The problem of the proper choice of $\partial^{-1}_x$ in (\ref{kp1evol})
requires a special investigation in the neighborhood of the initial time
$t=0$. The final answer is that, for initial data
$u(x,y)\in{\cal S}$ which are arbitrarily chosen and, therefore, not
necessarily subjected to the constraint
\beq \label{t0constraint}
\int\!dx\,u(x,y)=0\,,
\eeq
the function $u(t,x,y)$ reconstructed by solving the inverse spectral
problem for  (\ref{Schrodin}) evolves in time  according to the equation
\beq
\label{evolutionkp1}
u_t(t,x,y)-6u(t,x,y)u_x(t,x,y)+u_{xxx}(t,x,y)=3\!\int^{x}_{-t\infty}
\!\!dx'\,u_{yy}(t,x',y)\,.
\eeq
Notice that the possibility that the inversion of $\partial_{x}$ can depend on
$t$ was already mentioned in the literature for the
Davey--Stewartson equation~\cite{ASM}.
\item
The time derivative $u_t(t,x,y)$ has at $t=0$ different left and right
limits and the condition
\beq \label{constraint1}
\int\!dx\,u(t,x,y)=0
\eeq
is dynamically generated by the evolution equation at times $t\not=0$, i.e.
for these times we recover the result obtained in~\cite{ablvill}.
\item
The primitives
$\int_{\pm\infty}^{x}$ or their antisymmetric combination
${1\over 2}\!\left(\int_{-\infty}^{x}-\int_{x}^{+\infty}\right)$ cannot be
exchanged with the limit $t\to0$. In fact we have
\beqs
&&\lim_{t\to +0}\int_{-\infty}^{x}\!\!dx'\,u(t,x',y)=
-\lim_{t\to +0}\int_{x}^{+\infty}\!\!dx'\,u(t,x',y)=\nonumber\\
&&\lim_{t\to +0}{1\over
2}\!\left(\!\int_{-\infty}^{x}-\int_{x}^{+\infty}\right)
\!\!dx'\,u(t,x',y)=\!\int_{-\infty}^{x}\!\!dx'\,u(x',y)
\eeqs
and
\beqs
&&\lim_{t\to -0}\int_{-\infty}^{x}\!\!dx'\,u(t,x',y)=
-\lim_{t\to -0}\int_{x}^{+\infty}\!\!dx'\,u(t,x',y)=\nonumber\\
&&\lim_{t\to -0}{1\over
2}\!\left(\!\int_{-\infty}^{x}-\int_{x}^{+\infty}\right)
\!\!dx'\,u(t,x',y)=-\!\int_{x}^{+\infty}\!\!dx'\,u(x',y)\,.
\eeqs
Thus the two forms of writing down the evolution form of the
Kadomtsev--Petviashvili equation suggested in \cite{ablvill} and in~(\ref
{evolutionkp1}) are equivalent but only in the last formulation the
integration over $x$ and the limit $t\to0$ can be exchanged showing
explicitly the discontinuity at $t=0$.
\item
Conditions (`constraints') higher than (\ref{constraint1}) can be obtained.
But some special care is needed due to the asymptotic behavior of $u$ as
given in~(\ref {asympt1}). E.g. in the condition
\beq
\label{constraint2}
\int\!dx\,x\,u_y(t,x,y)=0\,, \qquad t\not=0\,,
\eeq
the $y$-derivative cannot be extracted from the integral. Higher
conditions are nonlinear and can be given by means of recursion relations.
\item
In spite of the dynamic condition (\ref{constraint1}) we have that at all
times
\beq \label{int}
\int\!dx\!\int\!dy\,u(t,x,y)=\int\!\int\!\!dxdy\,u(0,x,y).
\eeq
The integrations in the l.h.s.\ are performed in the order explicitly
indicated. Since no any constraint is imposed to the initial data this
quantity is not necessarily equal to zero and gives
a nontrivial integral of motion.
\item
The properties of the Jost solutions of the equation~(\ref {Schrodin})
essentially differs at $t=0$ and $t\not=0$. In particular, at $t=0$ the
coefficients of
the asymptotic $1/{\bf k}$--expansion, where ${\bf k}$ denoted the spectral
parameter,
depend on the sign of the imaginary part ${\bf k}_{\Im}$ (cf.~\cite{Manakov}
and~\cite{nsreshi}), while at $t\not=0$ these coefficients in the limit
$t\to0$
depend on the sign of $t$.
\item
Spectral data do not belong to the Schwartz space ${\cal S}$ and no any
small norm or smoothness condition can improve this situation.
\end{enumerate}

All along the article we find convenient to consider the problem in
terms of the Fourier transform of $u$
\beq
\label{fourier}
v(t,p)\equiv v(t,p_1,p_2)={1\over (2\pi)^{2}}\int\!\!\!\int\!\!dxdy\,
e^{i(p_1x-p_2y)}u(t,x,y)\,,
\eeq
where $p$ denotes a 2--component vector
\beq
\label{2comp}
p\equiv \{p_1,p_2\}\,.
\eeq
Then, the initial value problem~(\ref {kp1}) for the KPI equation can be
rewritten as $(d^2q\equiv dq_1dq_2)$
\beqs
p_1{\partial v(t,p)\over \partial t}&=&-3ip_{1}^2 \!\int\!\!d^{2}q\,v(t,q)
\,v(t,p-q)- i\,\bigl(p_{1}^{4}+3p_{2}^{2}\bigr)\,v(t,p)\,,\nonumber\\
v(0,p)&=&v(p)\in{\cal S}\,,
\label{kp1'}
\eeqs
where $v(p)$ is the Fourier transform~(\ref {fourier}) of the initial data
$u(x,y)$. The problem of properly defining $\partial_{x}^{-1}$
in~(\ref {kp1evol}) is reformulated as the problem of properly regularizing
the distribution $p_{1}^{-1}$.  In terms of the Fourier transform $v(t,p)$
we have
instead of~(\ref {evolutionkp1})
\beq
\label{kp2evol}
{\partial v(t,p)\over \partial t}=-3ip_{1} \!\int\!\!d^{2}q\,v(t,q)
\,v(t,p-q)- i\,{p_{1}^{4}+3p_{2}^{2}\over p_{1}+i0t}\,v(t,p)
\eeq
and condition~(\ref {constraint1}) takes the form
\beq
\label{constr1}
\lim_{p_1\to 0}v(t,p_1,p_2)=0,\qquad \mbox{for}\quad t\not=0,
\eeq
where the limit is understood in the sense of distributions in the $p_2$
variable.

According to the usual scheme of the spectral transform theory we expect
that the spectral data evolve in time as the Fourier transform of the linear
part of~(\ref{kp1}).
This makes clear that the special behavior in time of the
solutions of KPI is just complicated by the nonlinearity of the
KPI but is, in fact, inherent to the singular character of
its linear part. Due to this we start the article by considering in
section 2  the linearized version of the KPI equation. The above mentioned
properties are proved for the class of Schwartz initial
data. As well some more general classes of initial data are considered,
for which the condition of unique solvability of the initial value problem
is preserved and the corresponding evolution forms of the equation are
presented. The
study of the nonlinear case for these initial data is an open
problem.  It is show that in spite of the discontinuity at $t=0$ the
linearized evolution equation for initial data belonging to the Schwartz
space  is an Hamiltonian system. Thus it is natural to expect that the KPI
equation with this type of initial data is Hamiltonian too. But since a
formulation of the simplectic and Poisson structures in terms of spectral data
meets with some difficulties even in $1+1$ dimensions, i.e. in the KdV case
(see, e.g.~\cite{pbkdv}), we postponed this problem to a
forthcoming publication.

In section 3 the inverse spectral transform method for the KPI equation is
revisited. We use a specific form of the equations of the inverse
problem suggested in~\cite{total}. The advantages of this formulation as
well as the connections with the results obtained in~\cite{kpshort} are
given in
section 4. A lemma and some propositions on distributions that are of common
use in the article are reported in the appendix.


\section{The Linearized Equation}
\setcounter{equation}{0}

We consider the linear part of the KPI equation
\begin{equation}
\label{lin1}
\partial_{x} (\partial_{t} U(t,x,y)+\partial^{3}_{x} U(t,x,y))=
3\partial^{2}_{y}U(t,x,y).
\end{equation}
and its initial value problem, i.e. we are searching special classes of
solutions $U(t,x,y)$ satisfying
\begin{equation}
\label{initialU}
U(0,x,y)=U(x,y).
\end{equation}
The solution $U(t,x,y)$ and the initial value $U(x,y)$ can be distributions
in the $(t,x,y)$ variables belonging to the space $\cal S'$ dual to the
Schwartz space $\cal S$, but $U(t,x,y)$ must be continuous at $t=0$ in order
to have a meaningful initial value problem.

Since (\ref{lin1}) is not an evolution equation its initial value problem is
singular and we must pay special attention to the behaviour of the solution
$U(t,x,y)$ at the initial time $t=0$. From the study of this problem we
expect to get useful guidelines for the KPI equation. In fact, the
additional nonlinear term present in the KPI equation complicates the
analysis but does not change the specific behaviour of the solution in the
neighborhood to the initial time.

Performing the Fourier transform of the equation (\ref {lin1})
we get
\begin{equation} \label{lin1'}
p_{1}\,{\partial V(t,p)\over \partial t}=-i\,(p_{1}^{4}+3p_{2}^{2})
\, V(t,p)
\end{equation}
where
\beq
\label{fourier2}
V(t,p)\equiv V(t,p_1,p_2)={1\over (2\pi)^{2}}\int\!\!\!\int\!\!dxdy\,
e^{i(p_1x-p_2y)}U(t,x,y)
\eeq
and $p$ denotes a 2--component vector
\beq
\label{2comp2}
p\equiv \{p_1,p_2\}.
\eeq
For any $p_{1}\neq 0$ the general solution of~(\ref {lin1'}) satisfying the
 initial value requirement
\begin{equation}
\label{initialV}
V(0,p)=V(p)
\end{equation}
is given by
\begin{equation}
\label{1}
V(t,p)=\exp\left(-it\,{p_{1}^{4}+3p_{2}^{2}\over p_{1}}\right)\,V(p),
\end{equation}
We choose $V(p)$ to be the Fourier transform of the initial data $U(x,y)$
\begin{equation}
\label{fourierinitdata}
V(p) ={1\over(2\pi)^2}\!\int\!\!\int\!\!dx\,dy\,e^{ip_{1}x-ip_{2}y}U(x,y).
\end{equation}
Then (\ref{initialV}) is the Fourier transform of (\ref{initialU}) and the
initial value problem defined in (\ref{lin1}) and (\ref{initialU}) has been
transformed into an equivalent one in the Fourier transformed space of
variables $p$.

A natural regularization at $p_1=0$ of the distribution in
(\ref{1}) is just formula (\ref{1}) itself considered for all values of
the variables
$(t,p)$ including $p_1=0$. In order to
compute its time
derivative it is convenient to consider it as the limit in the sense of
$\cal S'$
\begin{equation}
\label{regularV}
V(t,p)\equiv \lim_{\epsilon\to+0}
\exp\left(-it\frac{p_1^4+3p_2^2}{p_1+i\epsilon t}\right)V(p).
\end{equation}
Then we derive with respect to $t$ both sides and exchange the time
derivative and the limit in the r.h.s., which is allowed in the space ${\cal
S}'$ since a test function multiplied by $\exp\left(-it\frac{p_1^4+
3p_2^2}{p_1+i\epsilon t}\right)$ still belongs to the Schwartz space. We get
(see lemma \ref{lemma1} in the appendix for an alternative proof) that $V(t,p)$
satisfies the evolution equation
\beq
\label{linevol2} {\partial V(t,p)\over \partial
t}=-i\,{p_{1}^{4}+3p_{2}^{2}\over p_{1}+i0t}\,V(t,p).
\eeq
Note that the special regularization of $1/p_1$ obtained in the r.h.s.\ is
fixed by the requirement that $V(t,p)$ evolves in time according to
(\ref{1}) for all values of $p_1$ including $p_1=0$.

The above considerations can be extended to the cases
in which
$\sgn p_1\,V(p)$, $p_2V(p)$ and $\frac{p_2}{|p_1|} V(p)$ belong to $\cal S$.
They need to be considered in details. For instance in the first case the
time derivative and the limit cannot be exchanged any more and the
time
derivative of $V(t,p)$ is not defined at $t=0$. A complete study of all
these cases is considered in the following.

We conclude that $V(t,p)$ furnishes via the Fourier transform
\begin{equation}
\label{Ureg}
U(t,x,y) = \!\int\!\! d^{2}p e^{-ip_{1}x+ip_{2}y}
V(t,p)
\end{equation}
a solution $U(t,x,y)$ belonging to the space of distributions
$\cal S'$ satisfying the initial data equation (\ref{initialU}) and
the evolution equation
\beq
\label{evolutionU}
\partial_tU(t,x,y)+\partial_{x}^3U(t,x,y)=
3\!\int^{x}_{-t\infty}
\!\!dx'\,\partial_{y}^2U(t,x',y).
\eeq
Note that this equation must be considered as an equation for distributions
in the variables $x$, $y$ and $t$. In particular at the initial time $t=0$
it can be continuous or discontinuous with left and right limits or not
defined at all. In the following sections we shall give an explicit example
for all these three possibilities.

Since the most general regularization of $V(t,p)$ at $p_1=0$ can be obtained
by adding a distribution with support the point
$p_1=0$ (see \cite{gelfsh}) we conclude that the general solution of the
problem~(\ref {lin1}), (\ref{initialU}) can be written in the form
\begin{equation}
\label{general}
U(t,x,y)+\Omega(t,x,y),
\end{equation}
where
\begin{equation}
\label{fourierOmega}
\Omega (t,x,y) = \!\int\!\! d^{2}p e^{-ip_{1}x+ip_{2}y}
\omega (t,p)
\end{equation}
is the Fourier transform of a distribution $\omega(t,p)$ concentrated at
$p_{1}=0$. We have
therefore to find the general solution of the problem
\begin{equation} \label{lin:omega}
p_{1}\,{\partial \omega (t,p)\over \partial t}=-i\,(p_{1}^{4}+3p_{2}^{2})
\, \omega (t,p),\qquad \omega (0,p)=0,
\end{equation}
where $\omega(t,p)$ is a finite sum of the $\delta$--function $\delta(p_1)$
and its derivatives of the form
\begin{equation}
\label{omega1}
\omega (t,p)=\sum_{n=0}^{N}\delta^{(n)}(p_{1})\,\omega_{n}(t,p_{2}),
\end{equation}
with $\omega _{n}$ some distributions in the $p_{2}$ and $t$ variables.
By inserting (\ref{omega1}) into (\ref {lin:omega}) and by using the known
properties
\begin{equation}
\label{pdelta}
p_{1}\,\delta(p_{1})=0,\qquad
p_{1}\,\delta^{(n)}(p_{1})=-n\delta ^{(n-1)}(p_{1}),\quad n=1,2,{\ldots}
\end{equation}
we get
\begin{equation}
\label{omega2}
p_{2}^{2}\omega_{N}(t,p_{2})=0.
\end{equation}
and, then, by recursion
\begin{equation}
\label{omega3}
\left(p_{2}^{2}\right)^{n+1}\omega_{N-n}(t,p_{2})=0\qquad n=1,{\ldots},N.
\end{equation}
Using again property~(\ref {pdelta}) for $p_{2}$ we get that
$\omega_{n}(t,p_{2})$ is a sum of the $\delta $-function $\delta(p_{2})$
and its derivatives up to the order $2(N-n)+1$ with time dependent
coefficients.

This result can be re-expressed in terms of the $x$ and $y$ variables by
saying that $\Omega(t,x,y)$ is a polynomial in $x$ and $y$ of the form
\begin{equation}
\label{Omega}
\Omega (t,x,y)=\sum_{n=0}^{N}\sum_{m=0}^{2(N-n)+1}\!a_{nm}(t)\,x^{n}\,y^{m}.
\end{equation}
The coefficients $a_{nm}(t)$ obey an undetermined linear system of
ordinary differential equations, which is obtained by inserting
$\Omega(t,x,y)$ into (\ref {lin1}).
It is easy to check (we omit for shortness the corresponding calculations)
that all the coefficients $a_{nm}(t)$
for $m=2,{\ldots},2(N-n)+1$ can be expressed as linear combinations of
$a_{n0}(t)$
and $a_{n1}(t)$ and their time derivatives. These functions are
arbitrary up to the condition that $\Omega(t,x,y)$ is identically zero at
the initial time $t=0$.

We conclude that
\begin{prop}\label{lin:reg}
For the initial data $U(x,y)$ we are considering, since they are vanishing
at large $x$ (or $y$), if there
exists a solution $U(t,x,y)$ of the linearized KPI
equation vanishing at large $x$ (or $y$) it is unique. This solution is
given by the
Fourier transform of the distribution $V(t,p)$ defined in
(\ref{1}) and satisfies the evolution equation (\ref{evolutionU}).
\end{prop}

In the following we will consider four different classes of initial data
vanishing at large $x$ or $y$ and we will show that in three cases the
Fourier transform of (\ref{1}) satisfies (in the sense of the distributions)
(\ref{evolutionU}) at all times
including $t=0$, vanishes at large $x$ or $y$ and is therefore unique, while
in one case the time derivative of the Fourier transform of (\ref{1})
at $t=0$ is not defined and therefore the evolution form of the linearized
KP equation has a (unique) solution in the space considered for any $t\not=0$.

\subsection{Some special classes of solutions of the linearized equation}

We want to build special classes of solutions of the evolution
version of the KPI equation in (\ref{evolutionU}). This can be done by
noting that, for any distribution  $G(t,x,y)$ satisfying the partial
differential equation
\begin{equation}
\label{Geq}
\left(\partial_t\partial_x-3\partial^2_y\right)G(t,x,y)=0,
\end{equation}
the convolution
\begin{equation}
\label{tau}
\tau(t,x,y)=\!\int\!\!\int\!\!dx'dy'G(t,x-x',y-y')U_0(t,x',y')
\end{equation}
of $G$ with an arbitrary solution $U_0$ of the equation
\begin{equation}
\label{U0eq}
\left(\partial_t+\partial^3_x\right)U_0(t,x,y)=0
\end{equation}
satisfies the linear partial differential equation
\begin{equation}
\label{taueq}
\left(\partial_t+\partial_x^3\right)\partial_x\tau=3\partial^2_y\tau
\end{equation}
and that this last equation can be considered the time evolution version of
(\ref{lin1}) for
\begin{equation}
\label{Ufromtau}
U(t,x,y)=\partial_x\tau(t,x,y).
\end{equation}
The general solution of (\ref{U0eq}) can be written as
\begin{equation}
\label{FTU0}
U_0(t,x,y)=\!\int\!\!\int\!d^2p\,e^{-ip_1x+ip_2y-itp_1^3}V_0(p)
\end{equation}
with $V_0(p)\in{\cal S'}$. We require the following integrals
\beqs
\label{intdxU0}
&&\int\!\!dx\,U_0(t,x,y)=\!\!\int\!\!dx\,U_0(0,x,y)= 2\pi\!\!\int\!\!dp_2\,
e^{ip_2y}V_0(0,p_2)\\
\label{intdyU0}
&&\int\!\!dy\,U_0(t,x,y)= 2\pi\!\!\int\!\!dp_1\,
e^{-ip_1x-itp_1^3}V_0(p_1,0)
\eeqs
to be well defined and convergent, i.e. we require $V_0(p)$ to be at $p_1=0$
and at $p_2=0$ a well defined distribution in the variable $p_2$ and $p_1$,
respectively.

Since we are studying an initial value problem special attention must be
paid to the distribution $G(t,x,y)$ and its derivatives at $t=0$. In
particular, since
\begin{equation}
U(0,x,y)=\!\int\!\!\int\!\!dx'dy'
{\left.\partial_xG(t,x-x',y-y')\right|}_{t=0}U_0(0,x',y')
\end{equation}
the distribution $G(t,x,y)$ must be defined at $t=0$.

Let us, now, consider the quadratic form
\begin{equation}
Q(t,x,y)=12xt-y^2
\end{equation}
and let us define
\beq
Q^{\lambda}_{-}=|Q|^{\lambda}\vartheta(-Q),\qquad
Q^{\lambda}_{+}=|Q|^{\lambda}\vartheta(Q).
\eeq
These distributions, following~\cite{gelfsh}, can be defined by analytical
continuation in $\lambda$. Both have simple poles at $\lambda=-1,-2,\dots,-k,
\dots$ and $Q^\lambda_+$ has additional simple poles at
$\lambda=-3/2,-3/2-1,\dots,-3/2-k,\dots$ The residua at the first kind of
poles are distributions with support on the cone surface $Q=0$ while the
residua at the second kind of poles are distributions with support on the
origin, i.e. on the vertex of the cone $Q=0$.

We are interested in using these distributions to build distributions
$G(t,x,y)$ satisfying the partial differential equation (\ref{Geq}). For all
the distributions introduced in the following we are using notations
and definitions of
\cite{gelfsh}. For instance the distribution ${1\over x}$ is defined by
\begin{equation}
\int \!\!dx\,{1\over x}\varphi(x)\equiv
\lim_{\epsilon\to0}\int_{|x|\geq\epsilon}\!\!dx\,{1\over x}\varphi(x)
\end{equation}
where $\varphi(x)$ is an arbitrary test function belonging to the Schwartz
space $\cal S$.

We are considering four cases:
\paragraph{Case i)}
\begin{equation}
G_+(t,x,y)=\frac{\sgn t}{\pi}Q_+^{-1/2}
\end{equation}
\paragraph{Case ii)}
\begin{equation}
\widehat{G}_{+}^{\sigma}(t,x,y)=\frac{\sgn t}{\pi}{y\over x}Q_+^{-1/2}
-2\sigma \delta(x)\vartheta(-\sigma y),\qquad \sigma=\pm
\end{equation}
\paragraph{Case iii)}
\begin{equation}
G_-(t,x,y)=\frac{1}{\pi}Q_-^{-1/2}
\end{equation}
\paragraph{Case iv)}
\begin{equation}
\widehat{G}_{-}(t,x,y)=\frac{1}{\pi}{y\over x}Q_-^{-1/2}.
\end{equation}
The corresponding solutions will be noted, respectively,
$U_+,{\widehat{U}}^{\sigma}_{+}, U_-,{\widehat{U}}_{-}$. The initial data
$U(0,x,y)$ can be computed in terms of the initial data $U_0(0,x,y)$ of the
evolution equation (\ref{U0eq}) that according to the requirements made for
the integrals in (\ref{intdxU0}) and (\ref{intdyU0}) must vanish at large
$x$ and $y$. It results (see in the following) that $U(0,x,y)$ vanishes at
large $x$ and does not necessarily vanish at large $y$. This is sufficient
to assure that all four cases furnish solutions belonging to the
special class considered in the previous section that admits one and only
one solution for the initial value problem.
\paragraph{Case i)}

It is convenient to exploit the general theorem about tempered distributions
assuming that any distribution can be expressed as derivative of a regular
function.

It is easy to verify that
\begin{equation}
\label{Q12+}
Q^{-1/2}_{+}-\pi\delta(y)\vartheta(xt)=\frac{\partial}{\partial y}\left[
\left(\arctan\frac{y}{{|Q|}^{1/2}}-\frac{\pi}{2}\sgn y\right)\vartheta(Q)
\right]
\end{equation}
and that
\begin{equation}
\label{Q12+additional}
Q^{-1/2}_{+}-\pi\delta(y)\vartheta(xt)=\frac{\partial^2}{\partial y^2}\left[
\left(y\arctan\frac{y}{{|Q|}^{1/2}}-\frac{\pi}{2}|y|\right)\vartheta(Q)
+Q_+^{1/2}\right]
\end{equation}
{}From (\ref{Q12+}) we have
\begin{equation}
\label{G+t0}
{\left.G_+(t,x,y)\right|}_{t=\pm0}=\pm\delta(y)\vartheta(\pm x)
\end{equation}
and from (\ref{Q12+additional}) by deriving with respect to $x$
\begin{equation}
\label{partialxQ}
\frac{\partial}{\partial x}Q_+^{-1/2}-\pi\sgn t\delta(x)\delta(y)=
{1\over2}\frac{\partial^2}{\partial y^2}\left[{1\over x}Q_+^{1/2}\right]
\end{equation}
and, consequently, at $t=0$
\begin{equation}
\label{partialxG+t0}
{\left.\frac{\partial}{\partial x}G_+(t,x,y)\right|}_{t=\pm0}=
\delta(x)\delta(y).
\end{equation}
{}From (\ref{partialxQ}) by deriving with respect to $t$ we get
\begin{equation}
\left(\frac{\partial}{\partial t}\frac{\partial}{\partial x}-
3\frac{\partial^2}{\partial y^2}\right)Q_+^{-1/2}=2\pi\delta(t)\delta(x)
\delta(y).
\end{equation}
{}From this equation, due to (\ref{partialxG+t0}), it follows that $G_+$
satisfies (\ref{Geq}).

We have therefore from (\ref{Ufromtau}) and (\ref{partialxG+t0})
\begin{equation}
U_+(0,x,y)=U_0(0,x,y)
\end{equation}
and from (\ref{tau}) and (\ref{G+t0})
\begin{equation}
\lim_{t\to\pm0}\tau_+(t,x,y)=\!\!\int_{\mp\infty}^x\!dx'U_0(0,x',y).
\end{equation}
\paragraph{Case ii)}

For studying the distribution $\widehat{G}_{+}^\sigma$ we consider the
identity
\begin{equation}
\label{G+tildedyy}
{y\over x}Q^{-1/2}_{+}\!\!+6\pi t\delta'(y)\vartheta(xt)=
-\frac{\partial^2}{\partial y^2}\!\left[6t\!
\left(\arctan\frac{y}{{|Q|}^{1/2}}-\frac{\pi}{2}\sgn y\right)\!\vartheta(Q)
+{1\over2}{y\over x}Q_+^{1/2}\!\right].
\end{equation}
{}From it at $t=0$ we derive
\begin{equation}
{\left.{y\over x}Q_+^{-1/2}\right|}_{t=0}=0
\end{equation}
and, consequently,
\begin{equation}
{\left.\widehat{G}^{\sigma}_{+}(t,x,y)\right|}_{t=0}=-2\sigma \delta(x)
\vartheta(-\sigma y)
\end{equation}
and
\begin{equation}
{\partial\over\partial x}
{\left.\widehat{G}^{\sigma}_{+}(t,x,y)\right|}_{t=0}=-2\sigma \delta'(x)
\vartheta(-\sigma y)
\end{equation}
By deriving (\ref{G+tildedyy}) with respect to $x$ we have
\begin{equation}
\frac{\partial}{\partial x}\left[{y\over x}Q^{-1/2}_{+}\right]+
6\pi |t|\delta'(y)\delta(x)=
{1\over2}\frac{\partial^2}{\partial y^2}\left[{y\over x^2}Q^{1/2}_{+}\right]
\end{equation}
and, then, by multiplying by $\sgn t$ and by deriving with respect to $t$
\begin{equation}
\left(\frac{\partial}{\partial t}\frac{\partial}{\partial x}-
3\frac{\partial^2}{\partial y^2}\right)\left[\sgn t{y\over x}Q_+^{-1/2}\right]
=-6\pi\delta'(y)\delta(x)
\end{equation}
and at $t=0$
\begin{equation}
\left.\frac{\partial}{\partial t}\frac{\partial}{\partial x}
\left[\sgn t{y\over x}Q_+^{-1/2}\right]\right|_{t=0}
=-6\pi\delta'(y)\delta(x).
\end{equation}
We deduce that $\widehat{G}^{\sigma}_{+}$ satisfies (\ref{Geq}).

We have therefore
\begin{equation}
{\widehat{U}}_{+}^{\sigma}(0,x,y)=
2\!\int^y_{\sigma\infty}\!dy'\partial_xU_0(0,x,y')
\end{equation}
and
\begin{equation}
\lim_{t\to\pm0}{\widehat{\tau}}_{+}^{\sigma}(t,x,y)=
2\!\int^y_{\sigma\infty}\!dy'U_0(0,x,y').
\end{equation}
Note that the initial data satisfy the constraint
\begin{equation}
\int\!\!dx\,{\widehat{U}}_{+}^{\sigma}(0,x,y)=0.
\end{equation}
\paragraph{Case iii)}

Let us, now, consider $Q_-^{-1/2}$. We have
\begin{equation}
\label{Q12-}
Q_-^{-1/2}=-\frac{\partial^2}{\partial y^2}\left[
y\ln\frac{\left|y-|Q|^{1/2}\right|}{|12xt|^{1/2}}
\vartheta(-Q)+Q_-^{1/2}\right]
\end{equation}
and by deriving it with respect to $x$
\begin{equation}
\frac{\partial}{\partial x}Q_-^{-1/2}=
-{1\over2}\frac{\partial^2}{\partial y^2}\left[{1\over x}Q_-^{1/2}\right]
\end{equation}
and at $t=0$
\begin{equation}
{\left.\frac{\partial}{\partial x}Q_-^{-1/2}\right|}_{t=0}=
-{1\over x}\delta(y)
\end{equation}
or
\begin{equation}
{\left.\frac{\partial}{\partial x}G_-(t,x,y)\right|}_{t=0}=
-{1\over\pi x}\delta(y).
\end{equation}
By deriving, in addition, with respect to $t$ we obtain that $G_-$ satisfies
(\ref{Geq}). However by considering (\ref{Q12-}) at $t=0$ it results that
$G_-$ is not defined at $t=0$ and, since $G_-$ satisfies (\ref{Geq}), that
also $\partial_t\partial_x G_-$ is not defined at $t=0$.

We have therefore
\begin{equation}
U_-(0,x,y)=-{1\over\pi}\!\int\!\!dx'\frac{1}{x-x'}U_0(0,x',y).
\end{equation}
However, the limit $\tau_-(t,x,y)$ for $t\to\pm0$ is not defined and the
evolution equation satisfied by $U_-(t,x,y)$ has no left and right limits at
$t=0$.
\paragraph{Case iv)}

Let us, now, consider the identity
\begin{equation}
\label{yoverxQ12-}
{y\over x}Q_-^{-1/2}=\frac{\partial^2}{\partial y^2}\left[
6t\ln\frac{\left|y-|Q|^{1/2}\right|}{|12xt|^{1/2}}
\vartheta(-Q)+{1\over2}{y\over x}Q_-^{1/2}\right].
\end{equation}
{}From it at $t=0$ we have
\begin{equation}
\label{Ghat-t=0}
{\left.\widehat{G}_{-}(t,x,y)\right|}_{t=0}={1\over\pi}\frac{\sgn y}{x}.
\end{equation}
and
\begin{equation}
{\partial\over\partial x}
{\left.\widehat{G}_{-}(t,x,y)\right|}_{t=0}= -{1\over\pi}\frac{\sgn y}{x^2}.
\end{equation}
By deriving (\ref{yoverxQ12-}) with respect to $t$ we get
\begin{equation}
\label{partialtQ12-}
\frac{\partial}{\partial t}\left[{y\over x}Q_-^{-1/2}\right]=
\frac{\partial^2}{\partial y^2}\left[
6\ln\frac{\left|y-|Q|^{1/2}\right|}{|12xt|^{1/2}}
\vartheta(-Q)\right]
\end{equation}
and, then, by deriving with respect to $x$ we deduce that
$\widehat{G}_{-}$ satisfies (\ref{Geq}).

We have therefore
\begin{equation}
{\widehat{U}}_{-}(0,x,y)=-{1\over\pi}\left(\int^y_{-\infty}+
\int^y_{\infty}\right) dy' \!\!\int\!\!dx'{1\over(x-x')^2}U_0(0,x,y)
\end{equation}
and
\begin{equation}
\lim_{t\to\pm0}{\widehat{\tau}}_{-}(t,x,y)={1\over\pi} \left(
\int^y_{-\infty}+
\int^y_{\infty}\right) dy'\!\!\int\!\!dx'{1\over x-x'}U_0(0,x,y).
\end{equation}
Note that the initial data satisfy the constraint
\begin{equation}
\int\!\!dx\,{\widehat{U}}_{-}(0,x,y)=0.
\end{equation}
In all four cases
\begin{equation}
\label{conditionU}
\int\!\!dx\,U(t,x,y)=0  \qquad \mbox{for}\,\, t\not=0.
\end{equation}
In the cases $ii)$ and $iv)$ this condition is satisfied also at $t=0$ and,
therefore, it can be considered as a constraint imposed to the initial data
that is conserved at all times. On the contrary in the cases $i)$ and $iii)$
the initial data do not satisfy (\ref{conditionU}), which can be considered
as a condition on the solution dynamically generated at all times $t\not=0$
by the evolution equation.

By inserting (\ref{FTU0}) into (\ref{tau}) we deduce that $\tau$ can be
rewritten as
\begin{equation}
\tau(t,x,y)=\!\!\int\!\!d^2p\,e^{-ip_1x+ip_2y-itp_1^3}H(t,p)V_0(p)
\end{equation}
where $H(t,p)$ is the Fourier transform of $G(t,x,y)$. The computation of
$H(t,p)$ allows us to state the connection with the general formula
(\ref{Ureg}) given in the previous section and, specifically, to state the
behaviour of $V(p)$ at $p_1=0$ and at $p_2=0$ in the four
cases considered via the formula
\begin{equation}
V(t,p)=-ip_1\,e^{-itp_1^3}H(t,p)V_0(p).
\end{equation}
It is convenient to compute $H(t,p)$  as a limit according to the formula
\begin{equation}
H(t,p)=\lim_{\mu\to0}\lim_{\epsilon_1\to0}\!\int\!\!dx(x^\mu_++x^\mu_-)
e^{ip_1x-\epsilon_1|x|}\lim_{\epsilon_2\to0}\!\int\!\!dy
e^{-ip_2y-\epsilon_2|y|}G(t,x,y).
\end{equation}
In fact the distribution $(x^\mu_++x^\mu_-)e^{-\epsilon_1|x|}
e^{-\epsilon_2|y|}G(t,x,y)$ converges to $G(t,x,y)$ as $\mu\to0$,
$\epsilon_1\to0$, $\epsilon_2\to0$ in the sense of distributions and the
Fourier transform operator commute with the limit. The insertion of the term
$x^\mu_++x^\mu_-$ is unessential in the case $G_\pm$ but in the cases
$\widehat{G}_{\pm}$ it takes into account explicitly that we have to compute
the principal value of the integration over $x$.

In all cases we make the change of variables $12x\sgn t\to x$ and
$y\to-|xt|^{1/2}y$. For the Bessel functions of different kinds introduced
in the following we use the notations of~\cite{EHI}.
\paragraph{Case i)}

For $G_+$ we get
\beqs
H_+(t,p)&=&\lim_{\epsilon_1\to0}\frac{\sgn t}{\pi}\!\int^\infty_0\!\!dx
e^{i({p_1\over12}\sgn t+i\epsilon_1)x}\!\!
\int^{+1}_{-1}\!\!dy(1-y^2)^{-1/2}e^{ip_2|xt|^{1/2}y}=\nonumber\\
&&\lim_{\epsilon_1\to0}\sgn t\!\int_0^\infty \!\!dx
e^{i({p_1\over12}\sgn t+i\epsilon_1)x}J_0\left(p_2|xt|^{1/2}\right)
\eeqs
and, therefore,
\begin{equation}
H_+(t,p)=\frac{i}{p_1+i0t}e^{-3it\frac{p_2^2}{p_1}}.
\end{equation}
\paragraph{Case ii)}

For $\widehat{G}^{\sigma}_{+}$ we get
\beqs
&&\int\!\!\int\!\!dx\,dy\,e^{ip_1x-ip_2y}{\sgn t\over\pi}{y\over x}Q_+^{-1/2}
(t,x,y)=\nonumber\\
&&\lim_{\epsilon_1\to0}-{|t|^{1/2}\over\pi}\! \int_0^\infty\frac{dx}{\sqrt{x}}
e^{i({p_1\over12}\sgn t+i\epsilon_1)x}\!\!
\int^{+1}_{-1}\!\!dy\,y(1-y^2)^{-1/2}e^{ip_2|xt|^{1/2}y}=\nonumber\\
&&\lim_{\epsilon_1\to0}i|t|^{1/2}\!\int_0^\infty\frac{dx}{\sqrt{x}}
e^{i({p_1\over12}\sgn t+i\epsilon_1)x}J_1(p_2|xt|^{1/2})=-{2i\over p_2}
\left(1-e^{-3it\frac{p_2^2}{p_1}}\right)
\eeqs
and, therefore,
\begin{equation}
\widehat{H}_{+}^{\sigma}(t,p)=-{2i\over
p_2+i\sigma0}e^{-3it\frac{p_2^2}{p_1}}.
\end{equation}
\paragraph{Case iii)}

For $G_-$ we get, analogously,
\begin{equation}
H_-(t,p)={1\over|p_1|}e^{-3it\frac{p_2^2}{p_1}}.
\end{equation}
\paragraph{Case iv)}

The computation of $\widehat{H}_{-}$ is much more involved and the insertion
of the term $x^\mu_++x^\mu_-$ becomes crucial. We obtain
\beqs
\label{Hhat-}
\widehat{H}_{-}(t,p)&=&\lim_{\mu\to0}\lim_{\epsilon_1\to0}-{\sgn t\over\pi}
|t|^{1/2}\left[\!\int_0^\infty\!\!dx\,x^{\mu-1/2}
e^{i({p_1\over12}\sgn t+i\epsilon_1)x}I^{(-)}(p_2,x,t)-
\right.\nonumber\\
&&\left.\!\int_0^\infty\!\!dx\,x^{\mu-1/2}
e^{-i({p_1\over12}\sgn t-i\epsilon_1)x}I^{(+)}(p_2,x,t)\right]
\eeqs
where
\beqs
I^{(-)}(p_2,x,t)&=&\lim_{\epsilon_2\to0}\!\!\int_1^\infty\!\!
dy\left(e^{i(p_2+i\epsilon_2)|xt|^{1/2}y}-
e^{-i(p_2-i\epsilon_2)|xt|^{1/2}y}\right)y(y^2-1)^{-1/2}=\nonumber\\
&&K_1\left(-i(p_2+i0)|xt|^{1/2}\right)-K_1\left(i(p_2-i0)|xt|^{1/2}\right)
\eeqs
and
\beqs
I^{(+)}(p_2,x,t)&=&\lim_{\epsilon_2\to0}\!\!\int_0^\infty\!\!
dy\left(e^{i(p_2+i\epsilon_2)|xt|^{1/2}y}-
e^{-i(p_2-i\epsilon_2)|xt|^{1/2}y}\right)y(y^2+1)^{-1/2}=\nonumber\\
&&-{i\over2}\left[K_1\left(-(p_2+i0)|xt|^{1/2}\right)
+K_1\left(-(p_2-i0)|xt|^{1/2}\right)-\right.\nonumber\\
&&\left.K_1\left((p_2-i0)|xt|^{1/2}\right)
-K_1\left((p_2+i0)|xt|^{1/2}\right)\right].
\eeqs
We need, then, to sum six integrals of the form
\begin{equation}
\label{intK1}
\int^\infty_0\!\!dx\,x^{\mu-1/2}e^{-\alpha x}K_1(2\beta\sqrt{x})=
\frac{\Gamma(\mu+1)\Gamma(\mu)}{2\beta}\alpha^{-\mu}\frac{\beta^2}{\alpha}
\Psi(\mu+1,2;\frac{\beta^2}{\alpha})
\end{equation}
extended to complex values of $\alpha$ and $\beta$. Note that (\ref{intK1})
at $\mu=0$, due to the presence of the gamma function $\Gamma(\mu)$, has a
simple
pole which must cancel out by summing up all terms in (\ref{Hhat-}).

In order to perform correctly the analytical continuation in $\alpha$
and $\beta$ and in order to extract the singular term at $\mu=0$ it is
convenient to rewrite the confluent hypergeometric function of second kind
$\Psi$ in terms of the confluent hypergeometric function $\Phi$ which is
an entire function by using the formula (see (63) at page 261 in \cite{EHI})
\beqs
&&\Gamma(\mu)\Gamma(\mu+1)\Psi(\mu+1,2;z)=\Gamma(\mu+1)\Phi(\mu+1,2;z)
\log z+\nonumber\\
&&\Gamma(\mu+1)\sum^\infty_{n=0}\frac{(\mu+1)_n}{(2)_n}\left[\psi(\mu+1+n)-
\psi(1+n)-\psi(2+n)\right]\frac{z^n}{n!}+\Gamma(\mu){1\over z}.
\eeqs
Therefore we get, for complex values of $\alpha$ and $\beta$,
\beqs
&&\int^\infty_0\!\!dx\,x^{\mu-1/2}e^{-\alpha x}K_1(2\beta\sqrt{x})=\nonumber\\
&&\frac{\Gamma(\mu+1)}{2\beta}\alpha^{-\mu}\frac{\beta^2}{\alpha}
\Phi(\mu+1,2;\frac{\beta^2}{\alpha})[2\log \beta-\log\alpha]+\nonumber\\
&&\frac{\Gamma(\mu+1)}{2\beta}\alpha^{-\mu}\frac{\beta^2}{\alpha}
\sum^\infty_{n=0}\frac{(\mu+1)_n}{(2)_n}\left[\psi(\mu+1+n)-
\psi(1+n)-\psi(2+n)\right]
\frac{{\left(\frac{\beta^2}{\alpha}\right)}^{n}}{n!}+\nonumber\\
&&\frac{\Gamma(\mu+1)}{2\beta}\alpha^{-\mu}.
\eeqs
By inserting it into the right hand side of (\ref{Hhat-}) we verify that the
coefficient of $\Gamma(\mu)$ at $\mu=0$ is zero. Therefore the limit $\mu\to0$
can be computed by using the Hospital rule. It is convenient to consider first
$\widehat{H}_{-}(t,p)$ for $p_1\not=0$. By recalling that
\begin{equation}
\Phi(1,2;z)={1\over z}(e^z-1)
\end{equation}
we get $(p_1\not=0)$
\begin{equation}
\widehat{H}_{-}(t,p)={2\over p_2}\sgn p_1 e^{-3it\frac{p_2^2}{p_1}}.
\end{equation}
The value of $\widehat{H}_{-}(t,p)$ at $p_1=0$ must be computed separately
and we get
\begin{equation}
{\left.\widehat{H}_{-}(t,p)\right|}_{p_1=0}=0.
\end{equation}

We get, therefore, for $\tau$
\paragraph{Case i)}
\begin{equation}
\tau_+(t,x,y)= \!\!\int\!\!d^2p\,e^{-ip_1x+ip_2y-itp_1^3-3i
t{p_2^2\over p_1} }\frac{i}{p_1+i0t} V_0(p)
\end{equation}
\paragraph{Case ii)}
\begin{equation}
{\widehat{\tau}}_{+}(t,x,y)=
-\!\!\int\!\!d^2p\,e^{-ip_1x+ip_2y-itp_1^3-3i
t{p_2^2\over p_1} }\frac{2i}{p_2+i\sigma 0} V_0(p)
\end{equation}
\paragraph{Case iii)}
\begin{equation}
\tau_-(t,x,y)= \!\!\int\!\!d^2p\,e^{-ip_1x+ip_2y-itp_1^3-3i
t{p_2^2\over p_1} }\frac{1}{|p_1|} V_0(p)\qquad (t\not=0)
\end{equation}
\paragraph{Case iv)}
\begin{equation}
{\widehat{\tau}}_{-}(t,x,y)=
\!\!\int\!\!d^2p\,e^{-ip_1x+ip_2y-itp_1^3-3it{p_2^2\over p_1}\,  }
\frac{2\sgn p_1}{p_2} V_0(p).
\end{equation}
We are, now, able to give more specific information on the general
distributional formula
\begin{equation}
\tau(t,x,y)=\!\!\int_{-t\infty}^x\!\!dx'\,\partial^2_yU(t,x',y)
\end{equation}
derived in the previous section. In the cases $ii)$ and $iv)$, since the
condition (\ref{conditionU}) is satisfied at all times including $t=0$, any
other operator $\partial^{-1}_x$ can be chosen instead of
$\int_{-t\infty}^xdx$,
and $\tau$ and, consequently, $\partial_tU(t,x,y)$ have a definite limit at
$t=0$. In the case $i)$ $\tau$ has definite right and left limit at $t=0$
and $\partial_tU(t,x,y)$ is in general discontinuous at $t=0$. In the case
$iii)$ $\tau$ is not defined at $t=0$ and at this point the distribution
$\partial_tU(t,x,y)$ is not regular.

For $V$ we get
\paragraph{Case i)}
\begin{equation}
V_{+}(t,p)=e^{-itp_1^3-3it{p_2^2\over p_1}\,  }V_0(p)
\end{equation}
\paragraph{Case ii)}
\begin{equation}
{\widehat{V}}_{+}^{\sigma}(t,p)=
-e^{-itp_1^3-3it{p_2^2\over p_1}\,  }\frac{2p_1}{p_2+i\sigma 0} V_0(p)
\end{equation}
\paragraph{Case iii)}
\begin{equation}
V_{-}(t,p)=-i\sgn p_1\,e^{-itp_1^3-3it{p_2^2\over p_1}\,  } V_0(p)
\end{equation}
\paragraph{Case iv)}
\begin{equation}
{\widehat{V}}_{-}(t,p)=
-i\,e^{-itp_1^3-3it{p_2^2\over p_1}\,  }
\frac{2|p_1|}{p_2} V_0(p).
\end{equation}
By taking into account that $V_0(p)$ is supposed to be continuous at $p_1=0$
and at $p_2=0$ the equations above show explicitly the behaviour of
$V(t,p)$ at these two special points in the four considered
cases.

\subsection{More on the case of smooth initial data}

We consider here, in more details, the case of initial data $U(x,y)$ (and
then $V(p)$) belonging to the test function Schwartz space ${\cal S}$ in the
$x$ and $y$ (correspondingly $p$) variables, i.e. the case $i)$ of the
previous section.

Let us recall that in this case the solution $U(t,x,y)$ of the linearized
KPI equation satisfies the
evolution equation
\beq \label{linevol4}
\partial
_{t}U(t,x,y)=-\partial ^{3}_{x}U(t,x,y)+3\partial ^{2}_y
\int^{x}_{-t\infty}dx'U(t,x',y)
\eeq
and can be written as
\begin{equation}
\label{UtJ}
U(t,x,y) =
\!\int\!\!\!\int\!\!  dx'dy'\,J(t,x-x',y-y')\,U_{0}(t,x',y')
\end{equation}
where we have introduced for convenience the notation
\begin{equation}
\label{explJ}
J(t,x,y)\equiv\partial_xG_{+}(t,x,y) = {\sgn t\over \pi }
\partial_{x}(12xt-y^{2})^{-1/2}_+ .
\end{equation}
The initial value of the evolution equation
\begin{equation}
\label{eqU0}
\partial_{t}U_{0}(t,x,y)=-\partial ^{3}_{x}U_{0}(t,x,y)
\end{equation}
satisfied by $U_0(t,x,y)$ coincides with the initial value of $U(t,x,y)$,
i.e.
\begin{equation}
\label{U0=U}
U_0(0,x,y)=U(x,y).
\end{equation}
Since, as we proved in the previous section,
\begin{equation}
\label{defJ}
J(t,x,y)=\!\int\!\!\!\int\!\!d^2 p\,
\exp\left(-ixp_1+iyp_2-3it\,{p_{2}^{2}\over p_{1}}\right)
\end{equation}
the solution can also be rewritten as
\beq
\label{Ut}
U(t,x,y)=\!\int\!\!d^{2}p\, e^{- i p_{1}x+i p_{2}y-i t p^{3}_{1}-3i
t {p ^{2}_{2}\over p_{1}} } V(p)
\eeq
where
\begin{equation}
\label{fourierinitdata'}
V(p) ={1\over(2\pi)^2}\!\int\!\!\int\!\!dx\,dy\,e^{ip_{1}x-ip_{2}y}U(x,y).
\end{equation}
The last formula for $U(t,x,y)$ can be used to get an integral of motion. In
fact by integrating (\ref{Ut}) first in $y$ and then in $x$ we obtain that
\beq
\label{integrallin}
\!\int\!\!dx\!\int\!\!dy\,U(t,x,y)=V(0)=\!\int\!\!\!\int\!\!dxdy\,U(x,y)
\eeq
for any $t$. This result does not contradict the condition
\beq
\label{constr}
\!\int\!\!dx\,U(t,x,y)=0,\qquad t\neq 0,
\eeq
that we proved, in the previous section, to be satisfied at any time
$t\not=0$ since the order of integration in (\ref{integrallin}) cannot be
exchanged.

By means of (\ref {UtJ}) and (\ref {explJ})  we  derive
that  $U(t,x,y)$, for $t$ ($t\neq 0$) and $y$ fixed, decreases rapidly
for $x\to -t\infty$ while for $x\to t\infty$ it
satisfies the following asymptotic behavior
\beqs
\label{asymptotics}
U(t,x,y)&= &{-1\over 4\pi \sqrt{3tx}|x|}\!\int\!\!\!\int\!\!dx'dy'\,U(x',y')\\
& &-{1\over 32\pi |t|\sqrt{3tx}x^2}\!\int\!\!\!\int\!\!dx'dy'
\,[12tx'+(y-y')^2]\,U(x',y')+o(|x|^{-5/2}).\nonumber
\eeqs
This asymptotic expansion is differentiable in $t$ and twice differentiable
in $y$ and obeys (\ref {lin1}). Therefore we can differentiate in $t$
the condition~(\ref {constr}) and use (\ref {linevol4}). This procedure
leads to the next condition
\beq \label{constr2'}
\!\int\!\!dx\,x\,U_{yy}(t,x,y)=0,\qquad t\neq 0.
\eeq
Thanks to  the
asymptotic behavior of $U(t,x,y)$ at large $x$ derived in (\ref
{asymptotics}) one $y$-derivative can be extracted from the
integral getting
\beq \label{constr2}
\!\int\!\!dx\,x\,U_{y}(t,x,y)=0,\qquad t\neq 0,
\eeq
but it is impossible to remove the second derivative since
$\int\!\!dx\,x\,U(t,x,y)$ is divergent. This procedure can be continued to
get an infinite set of dynamically generated conditions.

Equation (\ref {asymptotics}) explains the role of constraints,
i.e. conditions of the type~(\ref {constr}) and~(\ref {constr2})
imposed on the initial data $U(x,y)$, in the asymptotic behavior of
$U(t,x,y)$ at large $x$. Precisely, for each additional constraint that
is satisfied we get an additional $x^{-1}$ factor in the decreasing
law.

We want, now, to prove that the dynamical system described by the evolution
equation (\ref{linevol4}) is Hamiltonian.

We define the Poisson brackets of two  functionals $F$ and $G$ which are
differentiable with respect to the smooth initial data $U(x,y)$ as follows
\begin{equation}
\label{pb1:reg}
\{F,\,G\}=\!\int\!\!\!\int\!\! d\xi  d\eta\,{\delta F\over \delta
U(\xi,\eta)} {\partial \over \partial \xi }\,{\delta G\over \delta U(\xi
,\eta )},
\end{equation}
Then
\begin{equation}
\{U(x,y),\,U(x' ,y' )\}=\partial_x\delta(x-x' )\, \delta (y-y' )
\end{equation}
and
\begin{equation}
\label{pbV}
\{V(p),\,V(p' )\}= {-ip_{1}\over (2\pi )^{2}}\, \delta (p_{1}+p'_{1})\,
\delta (p_{2}+p'_{2}).
\end{equation}
Due to (\ref {eqU0}) and (\ref{U0=U}) we easily find that, at equal times,
also $U_0(t,x,y)$ are canonical since they satisfy the same Poisson brackets
\begin{equation}
\label{pb2:reg}
\{U_{0}(t,x,y),\,U_{0}(t,x' ,y' )\}=\partial_x\delta(x-x' )\,\delta (y-y' ).
\end{equation}
By considering $U(t,x,y)$ as given by~(\ref{Ut}) and by using (\ref{pbV}) we
have
\beqs
&&\{U(t,x,y),\,U(t,x' ,y' )\}=\nonumber\\
&&\qquad \qquad\!\int\!\!d^{2}p\!\int\!\!d^{2}p'\,
e^{- i p_{1}x+i p_{2}y-i t p^{3}_{1}-3i
t {p ^{2}_{2}\over p_{1}} } \,e^{- i{p'}_{1}x'+i{p'}_{2}y'-it {p'}^{3}_{1}-3i
t {{p'}^{2}_{2}\over {p'}_{1}} }\times \nonumber\\
&&\qquad \qquad {-ip_{1}\over (2\pi )^{2}}\, \delta
(p_{1}+p'_{1})\,\delta (p_{2}+p'_{2})=\nonumber\\
&&\qquad \qquad\partial_x\delta(x-x' )\, \delta (y-y' ).
\label{pb:regU}
\eeqs

Let us introduce the Hamiltonian
\begin{equation}
\label{H}
H= {1\over 2}\!\int\!\!\!\int\!\!dx dy\,(\partial _{x}U(t,x,y))^{2}
-{3\over 4}\int  dx'  \mid x' \mid  \!\int\!\! \!\int\!\!  dx
dy\bigl(\partial _{y}U(t,x,y)\bigr)\bigl(\partial _{y}U(t,x-x' ,y)\bigr) .
\end{equation}
Notice that the integration in $x'$ in the second term cannot be performed
before the integration in $x$ and $y$ since the corresponding integral does
not exist.

We prove the following
\begin{prop}\label{ham:regprop} \hfill
\begin{enumerate}
\item[\sl i)]
The Hamiltonian $H$ is an integral of motion.
\item[{\sl ii)}] The time evolution equation (\ref {linevol4})
is Hamiltonian, i.e.\ it can be rewritten as follows
\begin{equation}
\label{ham:reg}
\partial _{t}U(t,x,y) = \{U(t,x,y), H\}.
\end{equation}
\item[{\sl iii)}] The Poisson structure introduced in~(\ref {pb1:reg}) is not
continuous in $t$ at $t=0$
\begin{equation}
\lim_{t\rightarrow \pm 0}\{U(t,x,y), H\} \neq \{U(0,x,y), H\}\nonumber
\end{equation}
\end{enumerate}
\end{prop}
{\sl Proof}.

{\sl i)\/} Due to~(\ref {Ut}) definition~(\ref {H}) can be
rewritten as
\begin{equation}
\label{HV}
H= 2\pi^{2}\!\int\!\!d^{2}p\,\left(p_{1}^{2}+3p_{2}^{2}\,p_{1}^{-2}
\right)V(p)\,V(-p),
\end{equation}
which is explicitly time independent. Notice that $p_{1}^{-2}$ in this
expression, which has been obtained as the Fourier transform of $|x'|$, is
a distribution defined as follows (see \cite{gelfsh})
\begin{equation}
\label{princ}
\!\int\!\!dp_{1}\,p_{1}^{-2}\,\varphi (p_{1})\equiv \lim_{\epsilon\to0}
\!\int_{|p|\geq\epsilon}\!\!dp_{1}\, {\varphi (p_{1})-\varphi (0)\over
p_{1}^{2} },
\end{equation}
where $\varphi (p_{1})$ is an arbitrary test function belonging to the
Schwartz space $\cal S$.

{\sl ii)} We use~(\ref {pbV}) and (\ref {HV}) to write down
\begin{equation}
\label{V0H}
\{V(p), H\} = -i\,{p_{1}^{4}+3p_{2}^{2}\over p_{1}}\,V(p),
\end{equation}
where $1/p_{1}$ has to be understood in the sense of
the principal value. Now from (\ref {1}), by noting that due to the presence
of the exponential term for $t\not=0$
\begin{equation}
\lim_{p_1\to0}V(t,p_1,p_2)=0
\end{equation}
in the sense of distributions in the variable $p_2$, we get the evolution
equation (\ref{linevol2}) for any $t\neq 0$.

{\sl iii)} Considering the limit for
$t\to\pm 0$  we get by using the lemma \ref{lemma1} in the appendix
\begin{equation}
\lim_{t\rightarrow \pm 0}\{U(t,x,y), H\} = - \partial ^{3}_{x} U(0,x,y) +
3 \int^{x}_{\mp\infty}dx'\,\partial ^{2}_{y}U(0,x' ,y).\nonumber
\end{equation}
On the other side due to~(\ref {fourierinitdata})
\beqs
\{U(0,x,y), H\} &=& \{U(x,y), H\} =\nonumber\\
&&\!\int\!\!d^2p \,e^{-ip_{1}x+ip_{2}y}\,\{V(p),H\}=\nonumber\\
&-& \partial ^{3}_{x} U(0,x,y) + {3\over 2}\!\int\!\! dx' \sgn (x-x')
\partial ^{2}_{y} U(0,x',y),
\label{discP}
\eeqs
where in the last line we used~(\ref {V0H}).



\section{Nonlinear case}
\setcounter{equation}{0}

\subsection{Initial remarks and notations}
The analysis of the linearized version of the KPI equation, as performed in
the previous section, indicates clearly that it is convenient to study the
KPI equation in the space $p=\{p_1,p_2\}$ of the Fourier transform of the
solutions
\beq
\label{fourier3}
v(t,p)\equiv v(t,p_1,p_2)={1\over (2\pi)^{2}}\int\!\!\!\int\!\!dxdy\,
e^{i(p_1x-p_2y)}u(t,x,y).
\eeq
Correspondingly, also the nonstationary Schr\"odinger equation
\beq
\label{Schrodin3}
(-i\partial _{y}+\partial ^{2}_{x}-u(x,y))\Phi =0,
\eeq
which is the spectral equation associated to KPI, has to be stated in the
Fourier space of the variables $p=\{p_1,p_2\}$.

If $\Phi(x,y,{\bf k})$
is the Jost solution of (\ref {Schrodin3}) the corresponding Jost solution
in the $p-$space can be defined by first shifting
$\Phi(x,y,{\bf k})$ and then by taking the Fourier transform according to
the following formula
\begin{equation}
\label{jost}
\nu (p|{\bf k})={1\over(2\pi)^2}\!\int\!\!dx\,dy\,e^{ip_{1}x-ip_{2}y}
e^{i\hbox{\smallbf k}x-i\hbox{\smallbf k}^{2}y}\Phi(x,y,{\bf k})\,.
\end{equation}
The specific role played by the spectral parameter $\bf k$ is
emphasized by
separating it from the $p$ variable by a $|$ and a {\bf bold} font is used
to indicate that $\bf k$ belongs to the complex plane.

Then the spectral equation (\ref{Schrodin3}) takes the form
\begin{equation}
\label{eqJdiff}
\bigl[{\cal L}(p)-2p_1{\bf k}\bigr]\nu(p|{\bf k})=
\!\int\!\!d^2p'\,v(p-p')\,\nu(p'|{\bf k})\,,
\end{equation}
where
\begin{equation}
\label{notations}
{\cal L}(p)= p_2- p_1^2\,.
\end{equation}
The integral equation determining the Jost solution \cite{ZManakov}
has, in
the Fourier space, the form
\begin{equation} \label{eqJ}
\nu(p|{\bf k})=\delta^2(p)+ {\rho(p|{\bf k})\over {\cal
L}(p)-2p_1{\bf k}}\,,
\end{equation}
where
\begin{equation}
\label{eqJ'}
\rho(p|{\bf k})= \!\int\!\!d^2p'\,v(p-p')\,\nu(p'|{\bf k})
\end{equation}
and $\delta^2(p)\equiv \delta(p_1)\delta(p_2)$. The new function
\begin{equation}\label{defrho}
\rho(p|{\bf k})=[{\cal L}(p)-2p_1{\bf k}]\,\nu(p|{\bf k})
\end{equation}
will play a crucial role in the following. Let us introduce
the boundary values of $\nu $ and $\rho $ on the real axis of the $\bf
k$--plane
\begin{equation}
\label{boundary}
\nu^\pm(p|k)=\lim_{\hbox{\smallbf k}_{\Im}\to \pm 0}
\nu(p|{\bf k}),\qquad \rho^\pm(p|k)=\lim_{\hbox{\smallbf k}_{\Im}\to \pm 0}
\rho(p|{\bf k}),\qquad k={\bf k}_{\Re}
\end{equation}
and write
\begin{equation}
\label{eqJreal}
\nu^\pm(p|k)=\delta^2(p)+ {\rho^\pm(p|k)\over {\cal
L}(p)-2p_1k\mp i0p_1}.
\end{equation}
Then the spectral data are most naturally obtained (see \cite{nsreshi}) by
computing $\rho(p|k)$ at the special value of $k$ that vanishes the
denominator
\begin{equation}
\label{def:sd}
\rho^\pm \!\left(p\Bigl| {{\cal L}(p)\over
2p_1}\right)=\!\int\!\!d^2p'\,v(p-p')\,\nu^\pm \!\left(p'\Bigl|
{{\cal L}(p)\over  2p_1}\right)\,.
\end{equation}
We conclude that the variables $p=\{p_1,p_2\}$ can be considered as the
spectral
variables of the spectral data. For their introduction in the general scheme
of the resolvent approach see \cite{nsreshi,resreg,total} and for analogous
variables used in the KPII case see \cite{FokasSung}. They are related to
the more
familiar
spectral variables $\{\alpha,\beta\}$ (see
e.g.~\cite{ZManakov,FokasAblowitz,KP}) by the
formulae
\beqs
\label{subst}
\bmatrix{ll}
\alpha =\displaystyle{{\cal L}(-p)\over -2p_1}\,,&\qquad\beta
=\displaystyle{{\cal
L}(p)\over 2p_1} \\[10pt]
p_1=\alpha-\beta\,,&\qquad p_2=\alpha^2-\beta^2\,.
\ematrix
\eeqs
We denote the spectral data in the variables $\{\alpha,\beta\}$ by
$r^{\pm}(\alpha,\beta)$.
They are related to the previous ones by
\begin{equation}
\label{stand:sd}
r^{\pm}(\alpha,\beta)=-2\pi i\rho^\pm(\beta-\alpha,\beta^2-\alpha^2|\alpha)\,,
\qquad \rho^\pm \!\left(p\Bigl| {{\cal L}(p)\over  2p_1}\right)={i\over 2\pi}
r^\pm \!\left({{\cal L}(p)\over  2p_1}, {{\cal L}(-p)\over -2p_1}\right).
\end{equation}
Spectral data using the standard spectral variables $\{\alpha,\beta\}$
evolve in
time according to the formula \cite{KP}
\begin{equation}
\label{derSD2'3}
r^{\pm}(t,\alpha,\beta) =\exp\left(4it(\alpha^{3}-\beta^3)\right)\,
r^{\pm}(\alpha,\beta)\,,
\end{equation}
while the spectral data $\rho^\pm$ evolve in time as follows
\begin{equation}
\label{Sdt3}
\rho^{\pm} \Bigl(t,p\Bigl|{{\cal L}(p)\over 2p_1} \Bigr)=
\exp\!\left(-it\,{p_{1}^{4}+3p_{2}^{2}\over p_{1}} \right)\,
\rho^{\pm} \Bigl(p\Bigl|{{\cal L}(p)\over 2p_1} \Bigr)\,,
\end{equation}
i.e. in the same way as the linearized KPI equation in the Fourier
transformed space. In fact this is the deep reason of the privileged role
played by the $\{p_1,p_2\}$ variables in all the theory.

Note that the Jacobian of the transformation $\{\alpha,\beta\}\to
\{p_1,p_2\}$
\begin{equation}
{\partial (p_1,p_2)\over \partial (\alpha ,\beta)}=2|\alpha-\beta |
\end{equation}
is not bounded and not always different from zero. Therefore, the use of the
two sets of spectral variables can be not equivalent. In fact (see
\cite{kpshort}) the use of the
$\{\alpha,\beta\}$
variables requires to perform a subtraction in some formulae  of the direct
and inverse problem, while, as we will show in the following, this is no
more necessary if one uses the $\{p_1,p_2\}$ variables.

Let $\hbox{\bfcal R}^{\sigma}_{\pm}$ denotes the integral operator
with kernel
\begin{equation}
\label{defR}
{\cal R}^{\sigma}_{\pm}(\alpha,\beta)=\delta (\alpha-\beta)\mp
\theta \!\left(\mp\sigma (\alpha-\beta)\right)\,
r^{\sigma}(\alpha ,\beta )\,,\qquad \sigma =+,-\,.
\end{equation}
Then the characterization conditions for the spectral
data~\cite{KP} can be written as (cf.~\cite{nsreshi})
\beqs
\label{charR}
&&\hbox{\bfcal R}^{\sigma}_{\pm}\,\hbox{\bfcal R}^{-\sigma\,\dagger}_{\pm}=
\hbox{\bfcal I}\,,\qquad \sigma =+,-\,,\\
\label{indep}
&&\hbox{\bfcal R}^{\sigma}_{+}\,\hbox{\bfcal R}^{\sigma\,\dagger}_{+}=
\hbox{\bfcal R}^{\sigma}_{-}\,\hbox{\bfcal R}^{\sigma\,\dagger}_{-}\,.
\eeqs
If the integral equation in~(\ref{eqJ}), (\ref{eqJ'}) has no homogeneous
solutions, which we always suppose, the operator $\hbox{\bfcal
R}^{\sigma}_{\pm}$ satisfies also the equation
\begin{equation}\label{charR2}
\hbox{\bfcal R}^{-\sigma\,\dagger}_{\pm}\,\hbox{\bfcal R}^{\sigma}_{\pm}
=\hbox{\bfcal I}\,,\qquad \sigma =+,-\,,
\end{equation}
which is complementary to (\ref {charR}). Two other sets of spectral
data are used in literature, the so called unitary
$\hbox{\bfcal S}$-matrix
\begin{equation} \label{defS}
\hbox{\bfcal S}=\hbox{\bfcal R}^{\sigma\,\dagger}_{+}\,
\hbox{\bfcal R}^{-\sigma}_{-}\,,\qquad
\hbox{\bfcal S}\,\hbox{\bfcal S}^{\,\dagger}=\hbox{\bfcal I}=\hbox{\bfcal
S}^{\,\dagger}\,\hbox{\bfcal S}\,,\qquad\sigma =+,-,
\end{equation}
and the self-adjoint integral operator $\hbox{\bfcal F}^{\,\sigma}$
\begin{equation} \label{defF}
\hbox{\bfcal F}^{\,-\sigma}=\hbox{\bfcal R}^{\sigma}_{\pm}\,
\hbox{\bfcal R}^{\sigma\,\dagger}_{\pm}\,,\qquad
\hbox{\bfcal F}^{\,\sigma\,\dagger}=\hbox{\bfcal F}^{\,\sigma}\,,\qquad
\sigma =+,-\,,\qquad
\hbox{\bfcal F}^{\,\pm}\,\hbox{\bfcal F}^{\,\mp}=\hbox{\bfcal I}\,.
\end{equation}
Notice that, thanks to (\ref{indep}), the operator
$\hbox{\bfcal F}^{\,\sigma}$ as defined in~(\ref {defF}) is
independent on the signs $\pm$ of the constituent operators
$\hbox{\bfcal R}^{\sigma}_{\pm}$. Multiplying (\ref {indep}) by
$\hbox{\bfcal R}^{-\sigma\,\dagger}_{+}$ from the left and by
$\hbox{\bfcal R}^{-\sigma}_{-}$ from the right we get $\hbox{\bfcal
R}^{\sigma\,\dagger}_{+}\hbox{\bfcal R}^{-\sigma}_{-}=
\hbox{\bfcal R}^{-\sigma\,\dagger}_{+}\hbox{\bfcal R}^{\sigma}_{-}$, due
to~(\ref {charR2}). This proves the independence of the $\hbox{\bfcal
S}$-matrix
defined in~(\ref {defS}) on the sign $\sigma $. Properties of
$\hbox{\bfcal S}$ and $\hbox{\bfcal F}^{\,\sigma}$ given in~(\ref {defS}) and
(\ref {defF}) follow as well from~(\ref {charR}) and~(\ref {charR2}).
Taking into account the triangularity property of the operators
$\hbox{\bfcal R}^{\sigma}_{\pm}$ the six equations~(\ref
{charR}) and (\ref{indep}) reduce to the four ones~\cite{KP}
\beqs
\label{charR'}
&&r^{\sigma}(\alpha ,\beta)+\overline{r^{-\sigma}}(\beta,\alpha)=
\sigma \!\int^{\beta}_\alpha\!\!d\gamma\,
r^{\sigma}(\alpha,\gamma )\,\overline{r^{-\sigma}}(\beta,\gamma)\\
&&r^{\sigma}(\alpha ,\beta)+\overline{r^{\sigma}}(\beta,\alpha)=
\sigma \left(\int^{\infty}_\alpha-\int^\beta_{-\infty}\right)d\gamma\,
r^{\sigma}(\alpha,\gamma )\,\overline{r^{\sigma}}(\beta,\gamma)
\label{charforgotten}
\eeqs
and, analogously, instead of~(\ref {charR2}) we can write
\begin{equation}
\label{charR2'}
r^{\sigma}(\alpha ,\beta)+\overline{r^{-\sigma}}(\beta,\alpha)=
\sigma \!\int_\alpha^{\beta}\!\!d\gamma\,
r^{\sigma}(\gamma,\beta)\,\overline{r^{-\sigma}}(\gamma,\alpha)\,.
\end{equation}
Using now~(\ref {stand:sd}) we can rewrite (\ref{charR'}) and
(\ref{charforgotten}) in terms of $\rho$
\beqs
&&\rho^\sigma \!\left(p\Bigl| {{\cal L}(p)\over  2p_1}\right)-
\overline{\rho^{\sigma'}} \!\left(-p\Bigl| {{\cal L}(-p)\over -2p_1}\right)=
\nonumber\\
&&\qquad 2i\pi\sigma p_1\!\int\!\!d^{2}p'\,\sgn(p_1p'_1)
\vartheta \bigl(-\sigma\sigma'(p_1-p'_1)
p'_1\bigr)\,\delta \bigl(p_1{\cal L}(p')-p'_1{\cal L}(p)\bigr)\times
\nonumber
\\
&&\qquad \qquad  \rho^\sigma \!\left(p'\Bigl| {{\cal L}(p')\over
2p'_1}\right)\,\overline{\rho^{\sigma'}} \left(p'-p\Bigl| {{\cal
L}(p'-p)\over  2(p'_1-p_1)}\right)\label{char:rho}
\eeqs
Notice that due to the $\delta$--function and~(\ref {notations}) in the
integrand on the r.h.s.\ we have
\begin{equation}
\label{ident2}
{{\cal L}(p')\over  2p'_1}= {{\cal L}(p)\over 2p_1}\,,\qquad
{{\cal L}(p'-p)\over  2(p'_1-p_1)}= {{\cal L}(-p)\over  -2p_1}\,.
\end{equation}
Correspondingly from (\ref{charR2'}) we have
\beqs
&&\rho^\sigma \!\left(p\Bigl| {{\cal L}(p)\over 2p_1}\right)-
\overline{\rho^{-\sigma}} \!\left(-p\Bigl| {{\cal L}(-p)\over
-2p_1}\right)=\nonumber\\
&&\qquad \qquad 2i\pi\sigma p_1\!\int\!\!d^{2}p'\,\vartheta
\bigl((p_1-p'_1)p'_1
\bigr)\,\delta \bigl((p_1-p'_1){\cal L}(p)-p_1{\cal L}(p-p')\bigr)\times
\nonumber\\
&&\qquad \qquad \qquad \qquad \rho^\sigma \!\left(p'\Bigl| {{\cal
L}(p')\over  2p'_1}\right)\,
\overline{\rho^{-\sigma}} \!\left(p-p'\Bigl| {{\cal L}(p-p')
\over 2(p_1-p'_1)}\right)\,.\label{char:rho2}
\eeqs

\subsection{Properties of Jost solutions and Spectral Data}

\subsubsection{1/{\bf k}--expansion of the Jost solution at $t=0$}

To get the asymptotic $1/{\bf k}$--expansion of the Jost solution we
consider the
expansion of the kernel $({\cal L}(p)-2p_1{\bf k})^{-1}$ of the integral
equation~(\ref {eqJ}). By considering it as a distribution in the variables
${\cal L}(p)$ and $p_1$ we have that
\begin{equation} \label{expan}
{1\over {\cal L}(p)-2p_1{\bf k}}=-\sum_{n=0}^{\infty }{{\cal L}(p)^n
\over (2{\bf k})^{n+1}(p_1\pm i0{\cal L}(p))^{n+1}}\,,\quad {\bf k}\to \infty
,~~\pm {\bf k}_{\Im }>0.
\end{equation}
Note that the coefficients of the expansion
depend on the half-plane in which ${\bf k}$ tends to infinity.
Consequently, also the expansion of the Jost solution at large $\bf k$
depends on the half plane in which the limit is performed. If we note the
coefficients of the asymptotic expansion as follows
\begin{equation}
\label{decnu}
\nu(p|{\bf k})=\sum_{n=0}^{\infty}{\nu^{\pm}_{n}(p)\over (2{\bf k})^n},
\quad\rho(p|{\bf k})=\sum_{n=0}^{\infty}{\rho^{\pm}_{n}(p)\over
(2{\bf k})^n},\quad {\bf k}\to \infty
,~~\pm {\bf k}_{\Im }>0
\end{equation}
we have from (\ref {eqJ}) that
\beqs
&&\nu^{\pm}_{0}(p)=\delta ^2(p)\,,\qquad\rho^{\pm}_{0}(p)=v(p)\,,\nonumber\\
&&\nu^{\pm}_{n}(p)=-\sum_{m=0}^{n-1}{{\cal L}(p)^{n-m-1}\,
\rho^{\pm}_{m}(p)\over (p_1\pm i0{\cal L}(p))^{n-m}}\,,
\qquad n=1,2,{\ldots}\,,
\label{decnu2}
\eeqs
where $\rho^{\pm}_{n}(p)$ due to the~(\ref {eqJ'}) and (\ref {expan}) obey
the recursion relation
\begin{equation}\label{recur}
\rho^{\pm}_{n}(p)=-\sum_{m=0}^{n-1}\int\!\!{d^2q\,v(p-q)\,{\cal L}(q)^
{n-m-1}\over (q_1\pm i0{\cal L}(q))^{n-m}}\,\rho^{\pm}_{m}(q)\,,
\qquad n=1,2,{\ldots}\,.
\end{equation}
Only the
leading coefficients are independent on the sign of ${\bf k}_\Im$. The
coefficients with $n=1$
\begin{equation} \label{first}
\nu^{\pm}_{1}(p)=-{v(p)\over p_1\pm i0{\cal L}(p)}\,,\qquad
\rho^{\pm}_{1}(p)=-\!\int\!\!d^2q\,{v(p-q)\,v(q)\over q_1\pm
i0{\cal L}(q)}\,.
\end{equation}
will be of special use in the following.

\subsubsection{Properties of Spectral Data}

In order to study the properties of  the
spectral data we, first, substitute~(\ref{eqJ})
into~(\ref {eqJ'})
getting an integral equation for $\rho(p|{\bf k})$
\begin{equation} \label{eqrho}
\rho(p|{\bf k})=v(p)+ \!\int\!\!{d^2q\over {\cal
L}(q)-2q_1{\bf k}}\,v(p-q)\,\rho(q|{\bf k})\,.
\end{equation}
Performing the limiting procedure~(\ref{boundary}) we have
\begin{equation} \label{eqrhopm}
\rho^{\pm}(p| k)=v(p)+ \!\int\!\!{d^2q\over {\cal
L}(q)-2q_1 k\mp i0q_{1}}\,v(p-q)\,\rho^{\pm}(q| k)\,
\end{equation}
and, then,  recalling definition~(\ref {def:sd}), the following
representation for the spectral data
\begin{equation} \label{sd}
\rho^{\pm}\!\left(p\Bigl|{{\cal L}(p)\over  2p_1}\right)=
v(p)+ p_1\!\int\!\!{d^2q\over p_1{\cal
L}(q)-q_1{\cal L}(p) \mp i0p_{1}q_1}\,v(p-q)\,\rho^{\pm}\!\left(q\Bigl|
{{\cal L}(p)\over  2p_1}\right)\,.
\end{equation}
We deduce that for $v(p)$ vanishing rapidly at large $p$ also the spectral
data $\rho^{\pm}\!\left(p\Bigl|{{\cal L}(p)\over  2p_1}\right)$ vanish
rapidly  at large $p$. However, the spectral data are discontinuous at $p=0$.
In fact if the limit $p\to0$ is taken along the line $p_2=2\beta p_1$, with
$\beta$ an arbitrary constant,  we get from (\ref {sd})
\begin{equation} \label{sd0}
\lim_{p\to0\atop p_2=2\beta p_1} \rho^{\pm}\!\left(p\Bigl|{{\cal
L}(p)\over  2p_1}\right)=\rho^{\pm}(0|\beta)=
v(0)+\!\int\!\!{d^2q\,\overline{v(q)}\,\rho^{\pm}(q|\beta)
\over {\cal L}(q)-2\beta q_1 \mp i0q_1}\,,
\end{equation}
i.e. the limit depends on $\beta$.
In addition if we fix
$p_2\neq 0$ we get
\begin{equation} \label{sd0'}
\lim_{p_1\to 0}\rho^{\pm}
\!\left(p\Bigl|{{\cal L}(p)\over 2p_1}\right)=v(0,p_2)
\end{equation}
 and combining the results~(\ref {sd0}) and~(\ref {sd0'})
\begin{equation}
\label{sd0''}
\lim_{p_2\to 0}\lim_{p_1\to 0}\rho^{\pm}\!\left(p\Bigl|{{\cal L}(p)\over
2p_1}\right)=\lim_{\beta \to \infty}\lim_{p_1\to 0}\rho^{\pm}\!\left(p_1,
2\beta p_1\Bigl|\beta -{p_1\over 2}\right) =v(0)\,.
\end{equation}
Note that from (\ref{sd0'}) it results that the spectral data $\rho^\pm$ at
$p_1=0$  are
smooth (Schwartz) functions of $p_2$. Moreover, since
\begin{equation}
\label{rsd0'}
r^\pm(\beta,\beta)=-2\pi
i\rho^{\pm}(0|\beta)
\end{equation}
and
\begin{equation}
\label{rsd0''}
\lim_{\beta\to\infty}{i\over 2\pi} r^{\pm}\!\left(\beta -{p_2\over
4\beta},\beta+{p_2\over
4\beta} \right)
= \lim_{p_1\to 0}\rho^{\pm}
\!\left(p\Bigl|{{\cal L}(p)\over 2p_1}\right)\,,
\end{equation}
we deduce from (\ref{sd0}) and (\ref{sd0'}) that (no matter how
rapidly decreasing at large $p$ is chosen $v(p)$) the spectral data
$r^\pm(\alpha,\beta)$ do not vanish, in general, at large distances in the
$\{\alpha,\beta\}$ plane.
The spectral data
$r^\pm$ vanish at large distances  only if the constraint
\begin{equation}
v(0,p_2)=0
\end{equation}
is satisfied.

\subsubsection{1/{\bf k}--expansion of the Jost solution for $t\neq 0$}

Since the properties of the Jost solution depend on the specific form of the
spectral equation (\ref{eqJdiff}) we would expect that the introduction of a
parametric dependence on $t$ in $v(p)$ will not change the asymptotic
$1/{\bf k}$--expansion of $\nu(p|{\bf k})$ or, more precisely, we would
expect that the only difference would be a parametric dependence on $t$ of
the coefficients of the expansion. However, we will see that the time
evolution of a solution
$v(t,p)$ of the KPI equation has special properties in the neighborhoods of
$t=0$ and we have to take into account that
not $v(t,p)$ but $\widetilde{v}(t,p)$ as defined in the following equation
\begin{equation}
\label{nonzerov}
v(t,p)=e^{-it{p_{1}^{4}+3p_{2}^{2}\over p_{1}}}\,
\widetilde{v}(t,p)
\end{equation}
is continuous at $p=0$ for $t\not=0$.

Therefore in writing the integral equation (\ref{eqJ}) for
the Jost solution it is convenient to factorize explicitly the singular
behaviour
in time as follows
\begin{equation} \label{eqJt}
\nu(t,p|{\bf k})=\delta^2(p)+ {1\over {\cal
L}(p)-2p_1{\bf k}} \!\int\!\!d^2q\,e^{-it{q_{1}^{4}+3q_{2}^{2}\over
q_{1}}}\,\widetilde{v}(t,q)\,\nu(t,p-q|{\bf k})\,.
\end{equation}
We get
\begin{equation}
\lim_{\hbox{\smallbf k}\to\infty}{\bf k}\bigl[\nu (t,p|{\bf k})-\delta
^{2}(p)\bigr]=\widetilde{v}(t,p)\,\lim_{\hbox{\smallbf k}\to\infty}{{\bf
k}\over {\cal L}(p)-2p_1{\bf k}} \,e^{-it{p_{1}^{4}+3p_{2}^{2}\over
p_{1}} }\,,\nonumber
\end{equation}
and using proposition~\ref{lemma2} in the appendix
\begin{equation}
\label{1t}
\lim_{\hbox{\smallbf k}\to\infty}{\bf k}\bigl[\nu (t,p|{\bf k})-\delta
^{2}(p)\bigr]={-1\over 2(p_1+i0t)}\,e^{-it{p_{1}^{4}+3p_{2}^{2}\over
        p_{1}}} \widetilde{v}(t,p)={-1\over 2}\,{v(t,p)\over p_1+i0t} \,.
\end{equation}
Comparing this result with~(\ref{first}) we see that the
coefficients of $1/{\bf k}$ in the asymptotic expansions coincide only if
\begin{equation}
\label{sign1}
\sgn {\bf k}_{\Im}=\sgn(t{\cal L}(p))\,,
\end{equation}
or in other words only under this condition the limits ${\bf k}\to\infty$ and
$t\to 0$ commute. Notice that this condition is formulated in terms of the
variables $p$ corresponding to the fact that the special behaviour of
solutions and Jost solutions in the neighborhoods of $t=0$ are manageable
only in the transformed Fourier space.


\subsection{Inverse problem and time evolution}


\subsubsection{Formulation of the inverse problem}

By working in the general framework of the resolvent approach
\cite{resreg,total} we obtained in \cite{nsreshi} the following coupled
system of two linear integral equations for the boundary values of the Jost
solutions on the real $k$--axis
\begin{equation} \label{IP}
\nu^\pm(p|k)=\delta^2(p)+{1\over 2} \!\int\!\!{d^2q\over {\cal
L}(-q)+2kq_1\pm iq_10}\,\sum
_{\sigma=+,-}\overline{\rho^{-\sigma}}\Bigl(q\Bigl|{{\cal L} (q)\over
2q_1}\Bigr)\,\nu^{\sigma}\Bigl(p+q\Bigl|{{\cal L}(q)\over 2q_1}\Bigr)\,.
\end{equation}
This formula is the main tool to be used for solving the inverse problem,
i.e. for reconstructing the potential $v(p)$ starting from the spectral data
and for proving that spectral data evolving in time as indicated in
(\ref{Sdt3}) generate potentials $v(t,p)$ that are solutions of the KPI
equation.

In this respect two properties of this system are crucial. First, the system
is closed in the sense that the spectral data $\rho^\pm$ are reductions of
the Jost solutions themselves via equations (\ref{defrho}), (\ref{boundary})
and (\ref{def:sd}). Second, it depends linearly on the spectral data, in
contrast with the usual equations written in the literature that are
quadratic in the spectral data. To our knowledge the only alternative
integral equations for solving the inverse problem, which are linear in the
spectral data, are given in \cite{Zhou}, but they are not closed in the sense
that they are written in terms of the so called advanced/retarded solutions.

We suppose that the system (\ref{IP}) has unique solution in the
considered class of spectral data. Then the r.h.s.\ of~(\ref {IP})
can be continued analytically in the complex ${\bf k}$-plane as follows
\begin{equation} \label{IPcompl}
\nu(p|{\bf k})=\delta ^2(p)+{1\over 2} \!\int\!\!{d^2q\over {\cal L}(-q)+
2{\bf k}q_1}\,\sum _{\sigma =\pm}\overline{\rho^{-\sigma}}
\Bigl(q\Bigl|{{\cal L}(q)\over 2q_1}\Bigr)\,\nu^{\sigma}\Bigl(p+q\Bigl|
{{\cal L}(q)\over 2q_1}\Bigr)\,.
\end{equation}
For getting from (\ref{IPcompl}) an integral equation for $\rho(p,{\bf k})$
it is convenient to use equation (\ref {eqJ}) computed on the real
axis and
to rewrite the Jost solution in the r.h.s.\ of (\ref{IPcompl}) as
\begin{equation} \label{1'}
\nu^{\sigma}\Bigl(p+q\Bigl|{{\cal L}(q)\over 2q_1}\Bigr)=
\delta^2 (q+p)+{q_1\,\rho^{\sigma}\bigl(p+q\bigl|{{\cal L}(q)\over
2q_1}\bigr)\over q_1{\cal L}(p)+p_1{\cal L}(-q)-i\sigma 0q_1(q_1+p_1)}\,.
\end{equation}
Then it is easy to check that
\begin{equation} \label{2}
{\nu^{\sigma}\bigl(p+q\bigl|{{\cal L}(q)\over
2q_1}\bigr)\over {\cal L}(-q)+2{\bf k}q_1}={1\over
{\cal L}(p)-2{\bf k}p_1}\left[-{p_1\over q_1}\,\nu^{\sigma}\Bigl(p+q\Bigl|
{{\cal L} (q)\over 2q_1}\Bigr)
+{\rho^{\sigma}\bigl(p+q\bigl|{{\cal L}(q)\over 2q_1}\bigr)
\over {\cal L}(-q)+2{\bf k}q_1}\right], \end{equation}
where the ratio $p_1/q_1$ multiplying $\nu$ in the r.h.s.\ is well
defined due to (\ref {1'}). By inserting this expression in (\ref{IPcompl})
and by recalling definition (\ref{defrho}) we get
the following integral equation for $\rho(p|{\bf k})$
\beqs
&&\rho(p|{\bf k})=-\!\int\!\!d^2q\,{p_1\over 2q_1} \,\sum _
{\sigma =\pm}\overline{\rho^{-\sigma}}\Bigl(q\Bigl|{{\cal L}(q)\over
2q_1}\Bigr)\,
\nu^{\sigma}\Bigl(p+q\Bigl|{{\cal L}(q)\over 2q_1}\Bigr)+\nonumber\\
&&\qquad{1\over 2} \int\!\!{d^2q\over {\cal L}(-q)+2{\bf k}q_1}\,\sum
_{\sigma =\pm}\overline{\rho^{-\sigma}}\Bigl(q\Bigl|{{\cal L}(q)\over
2q_1}\Bigr)\,
\rho^{\sigma}\Bigl(p+q\Bigl|{{\cal L}(q)\over 2q_1}\Bigr).\label{rho1}
\eeqs
It has a kernel similar to the kernel of the integral equation
(\ref{eqJ}), (\ref{eqJ'}) obtained for $\nu(p|{\bf k})$ in the direct problem.
Therefore its asymptotic expansion at large $\bf k$ can be computed by using
the same methods used in the previous section. In particular we get that the
first term in the r.h.s.\
due to~(\ref {decnu}) and (\ref{first}) is just
the potential
\begin{equation} \label{v1}
v(p)=-\!\int\!\!d^2q\,{p_1\over 2q_1}\,\sum_{\sigma =\pm}\overline{
\rho^{-\sigma}} \Bigl(q\Bigl|{{\cal L}(q)\over 2q_1}\Bigr)\, \nu^{\sigma}
\Bigl(p+q\Bigl|{{\cal L}(q)\over 2q_1}\Bigr)\,,
\end{equation}
that is therefore explicitly reconstructed in terms of the spectral data
and Jost solutions.

It is instructive to consider in details the properties of the function
$v(p)$ defined by  (\ref{v1}) in the neighborhoods of $p=0$. This analysis
throws some light on the meaning of the characterization equations for
spectral data reported in section 3.2. In fact, if we rewrite~(\ref
{v1}) by using (\ref {1'}) as
\beqs
v(p)&=&{1\over 2} \sum_{\sigma =\pm}\overline{\rho^{-\sigma}}
\Bigl(-p\Bigl|{{\cal L}(-p)\over -2p_1}\Bigr)-\nonumber\\
&&{p_1\over 2}\!\int\!\!d^2q\,\sum_{\sigma =\pm}{\overline{\rho^{-\sigma}}
\!\left(q\bigl|{{\cal L}(q)\over 2q_1}\right)\,\rho^{\sigma}
\!\left(p+q\bigl|{{\cal L}(q)\over 2q_1}\right)\over
q_1{\cal L}(p)+p_1{\cal L}(-q)-i\sigma 0q_1(q_1+p_1)}\,.
\label{v2}
\eeqs
and perform the same limit as in (\ref{sd0}) we get
\beqs
\lim_{p_1\to \pm 0}v(p_1,2\beta p_1)&=&
{1\over 2} \sum_{\sigma
=\pm}\Biggl\{\overline{\rho^{-\sigma}}(0|\beta)-\nonumber\\
&&\int\!\!d^2q\,{\overline{\rho^{-\sigma}}
\!\left(q\bigl|{{\cal L}(q)\over 2q_1}\right)\,\rho^{\sigma}
\!\left(q\bigl|{{\cal L}(q)\over 2q_1}\right)\over
{\cal L}(-q)+2\beta q_1\mp i\sigma 0}\Biggl\}.
\label{v3}
\eeqs
This result seems to contradict the fact that the potential $v(p)$ was
chosen to be continuous at $p=0$. However, the term in the r.h.s
which depends on the sign of $p_1$ is proportional to $\sum_{\sigma
=\pm}\sigma \!\int\!\!d^2q\,
\delta ({\cal L}(-q)+2\beta q_1)\bigl(\overline{\rho^{-\sigma}}\rho^{\sigma}
\bigr)\!\left(q\bigl|{{\cal L}(q)\over 2q_1}\right)$ and, then,
due to~(\ref {stand:sd}) and (\ref{defR})  to
$\sum_{\sigma =\pm}\sigma \!\int\!\!d\alpha \overline{r}^{\,-\sigma}
(\alpha,\beta)r^{\sigma}(\alpha,\beta)=
\sum_{\sigma =\pm}\sigma \Bigl(\bigl(\hbox{\bfcal R}^{-\sigma\,\dagger}_{+}-
\hbox{\bfcal R}^{-\sigma\,\dagger}_{-}\bigr)\bigl(\hbox{\bfcal R}^{\sigma}
_{+}-\hbox{\bfcal R}^{\sigma}_{-}\bigr)\Bigr)(\beta,\beta)$. But due
to~(\ref {charR2}) and~(\ref {defS})
$\bigl(\hbox{\bfcal R}^{-\sigma\,\dagger}_{+}-
\hbox{\bfcal R}^{-\sigma\,\dagger}_{-}\bigr)\bigl(\hbox{\bfcal R}^{\sigma}
_{+}-\hbox{\bfcal R}^{\sigma}_{-}\bigr)=2\hbox{\bfcal I}-
\hbox{\bfcal S}-\hbox{\bfcal S}^{\dagger}$ which is $\sigma$-independent.
Therefore this term is equal to zero and the limit does not depend on the
sign of $p_1$. The independence on $\beta$ can be proved by using (\ref{sd0}).

{}From (\ref {IPcompl}) by using the expansion (\ref{expan}) we can get the
coefficients of the asymptotic expansion at large $\bf k$ of $\nu(p|{\bf k})$.
Of special interest is the coefficients of $1/{\bf k}$. Recalling (\ref
{first}) we have
\begin{equation} \label{v4}
{v(p)\over p_1\pm i0{\cal L}(p)} =-\!\int\!\!{d^2q\,\over
2(q_1\mp i0{\cal L}(-q))}\,\sum_{\sigma =\pm}\overline{\rho^{-\sigma}}
\Bigl(q\Bigl|{{\cal L}(q)\over 2q_1}\Bigr)\,
\nu^{\sigma}\Bigl(p+q\Bigl|{{\cal L}(q)\over 2q_1}\Bigr)\,.
\end{equation}
For the all set of coefficients of the asymptotic expansion of $\rho(p|{\bf
k})$ we get
\begin{equation} \label{coeff1}
\rho^{\pm}_{n}(p)=\!\int\!\!{d^2q\,(-{\cal L}(-q))^n\over 2(q_1\mp
i0{\cal L}(-q))^{n+1}}\,\sum_{\sigma=\pm}\overline{\rho^{-\sigma}}
\Bigl(q\Bigl|{{\cal L}(q)\over 2q_1}\Bigr)\,\rho^{\sigma}
\Bigl(p+q\Bigl|{{\cal L}(q)\over 2q_1}\Bigr)\,.
\end{equation}

\subsubsection{Jost solutions for $t\neq 0$}

We choose spectral data to evolve as in (\ref{Sdt3})
\begin{equation}
\label{Sdt}
\rho^{\pm} \Bigl(t,p\Bigl|{{\cal L}(p)\over 2p_1} \Bigr)=
\exp\!\left(-it\,{p_{1}^{4}+3p_{2}^{2}\over p_{1}} \right)\,
\rho^{\pm} \Bigl(p\Bigl|{{\cal L}(p)\over 2p_1} \Bigr), \end{equation}
where $\rho^{\pm}(p|{{\cal L}(p)\over 2p_1})$ denotes the initial
value of spectral data determined in terms of $v(p)$ and Jost solutions
$\nu^\pm(p|k)$ according to formula (\ref{def:sd}).

Then the Jost solution at any time is reconstructed just by inserting
these spectral data into (\ref{IPcompl})
\begin{equation} \label{IPcompl:t}
\nu(t,p|{\bf k})=\delta ^2(p)+{1\over 2} \!\int\!\!{d^2q\,
e^{it{q_{1}^{4}+3q_{2}^{2}\over q_{1}}}
\over {\cal
L}(-q)+2{\bf k}q_1}\,\sum _{\sigma =\pm}\overline{\rho^{-\sigma}}
\Bigl(q\Bigl|{{\cal L}(q)\over 2q_1}\Bigr)\,\nu^{\sigma}\Bigl(t,p+q\Bigl|
{{\cal L}(q)\over 2q_1}\Bigr).
\end{equation}
In the limit ${\bf k}\to k\pm i0$ (cf.~(\ref {IP})) we obtain the integral
equations for the boundary values on the real axis, i.e. for the functions
$\nu^{\pm}(t,p|k)$. In what follows we always suppose that these equations
are uniquely solvable.

We have, first, to prove that this function $\nu(t,p|{\bf k})$ is indeed the
Fourier
transformed of the
Jost solution of (\ref
{Schrodin}) , i.e. that it satisfies (\ref{eqJdiff}), and, second, we have
to find
its time evolution equation, which is needed to know in order to show that
the reconstructed $v(t,p)$ satisfies KPI.

\subsubsection{The recontructed spectral equation}

First we prove that $\nu$ defined by~(\ref {IPcompl:t}) obeys the spectral
equation~(\ref {eqJdiff}) at any time. Let us define $\rho(t,p|{\bf k})$
by~(\ref{defrho}) rewritten for any time
\begin{equation}\label{defrho:t}
\rho(t,p|{\bf k})=[{\cal L}(p)-2p_1{\bf k}]\,\nu(t,p|{\bf k}).
\end{equation}
Then, by the same procedure used to get (\ref{rho1}) from
(\ref{IPcompl}) we derive from (\ref{IPcompl:t}) that
\begin{equation} \label{IP:rho}
\rho(t,p|{\bf k})=v(t,p)+{1\over 2} \!\int\!\!
{d^2q\,e^{it{q_{1}^{4}+3q_{2}^{2}\over q_{1}}}\over {\cal L}(-q)+
2{\bf k}q_1}\,\sum _{\sigma =\pm}\overline{\rho^{-\sigma}}
\Bigl(q\Bigl|{{\cal L}(q)\over 2q_1}\Bigr)\,\rho^{\sigma}\Bigl(t,p+q\Bigl|
{{\cal L}(q)\over 2q_1}\Bigr)\,,
\end{equation}
where we have introduced
\begin{equation} \label{v1t}
v(t,p)=-\!\int\!\!d^2q\,{p_1\over 2q_1}\,
\,e^{it{q_{1}^{4}+3q_{2}^{2}\over q_{1}}}\,
\sum_{\sigma =\pm}\overline{\rho^{-\sigma}}
\Bigl(q\Bigl|{{\cal L}(q)\over 2q_1}\Bigr)\,\nu^{\sigma}
\Bigl(t,p+q\Bigl|{{\cal L}(q)\over 2q_1}\Bigr). \end{equation}
Now due to the assumption on unique solvability of~(\ref
{IPcompl:t}) we have from~(\ref {IP:rho})
that $\rho(t,p|{\bf k})= \!\int\!\!d^2qv(t,p-q)\nu(t,q|{\bf k})$, i.e. (\ref
{eqJ'}) is satisfied for any $t$. Thanks to~(\ref {defrho}) this
proves that (\ref {eqJdiff}) is satisfied at any time.

To prove that $\nu$ defined by~(\ref {IPcompl:t}) obeys the integral
equation~(\ref {eqJ}) at any time we use (\ref {IPcompl:t}), (\ref {IP:rho})
and (\ref
{v1t}) to get
\beqs
&&\nu(t,p|{\bf k})-\delta^2(p)- {\rho(t,p|{\bf k})\over {\cal L}(p)-2p_1{\bf
k}}=\nonumber\\
&&{1\over 2} \!\int\!\!{d^2q\,e^{it{q_{1}^{4}+3q_{2}^{2}\over q_{1}}}
\over {\cal L}(-q)+2{\bf k}q_1}\,\sum _{\sigma =\pm}\overline{\rho^{-\sigma}}
\Bigl(q\Bigl|{{\cal L}(q)\over 2q_1}\Bigr)\times\nonumber\\
&&\left[\nu^{\sigma}\Bigl(t,p+q\Bigl|{{\cal L}(q)\over 2q_1}\Bigr)
\frac{{\cal L}(p+q)-2(p_1+q_1){{\cal L}(q)\over 2q_1}}{{\cal L}(p)-2
{\bf k}p_1}-\frac{\rho^{\sigma}\bigl(t,p+q\bigl|{{\cal L}(q)\over 2q_1}}%
{{\cal L}(p)-2{\bf k}p_1}\right]\,.
\eeqs
Then~(\ref {eqJ}) at any $t$ follows by using definition
(\ref{defrho:t}) of $\rho^\sigma$ in the r.h.s. This proves that
$\nu(t,p|{\bf k})$ in
(\ref{IPcompl:t}) is the Fourier
transform of the Jost solution of the equation~(\ref {Schrodin}) with
potential the Fourier transform of $v(t,p)$ defined in (\ref{v1t}).

To prove that $v(t,p)$ satisfies equation (\ref{nonzerov}) with
$\widetilde{v}(t,p)$ continuous at $p=0$ it is convenient to insert
(\ref{1'}) (that we proved to be satisfied at any time) into (\ref{v1t})
getting
\beqs
v(t,p)&=&e^{-it{p_{1}^{4}+3p_{2}^{2}\over p_{1}}}\,
{1\over 2} \sum_{\sigma =\pm}\overline{\rho^{-\sigma}}
\Bigl(-p\Bigl|{{\cal L}(-p)\over -2p_1}\Bigr)-\nonumber\\
&&{1\over 2}\!\int\!\!d^2q\,p_1e^{it{q_{1}^{4}+3q_{2}^{2}\over q_{1}}}\,
\sum_{\sigma =\pm}{\overline{\rho^{-\sigma}}
\!\left(q\bigl|{{\cal L}(q)\over 2q_1}\right)\,\rho^{\sigma}
\!\left(t,p+q\bigl|{{\cal L}(q)\over 2q_1}\right)\over
q_1{\cal L}(p)+p_1{\cal L}(-q)-i\sigma 0q_1(q_1+p_1)}\,.
\label{v2t}
\eeqs
The time evolution of $\rho^{\sigma}\!\left(t,p+q\bigl|{{\cal L}(q)\over
2q_1}\right)$ in the r.h.s.\ is not explicitly known. However, it is useful
to factorize the time evolution it would have in the limit $v(p)\to0$ by
writing (see also (\ref{dernu}) below)
\begin{equation}
\rho(t,p|{\bf k})=e^{-2it[3({\bf k}+p_1){\cal
L}(p)+2p_1^3]}\,\widetilde{\rho}(t,p|{\bf k})\,.
\end{equation}
Note that for the special reduction furnishing the spectral data we have that
\begin{equation}
\label{rhotilde}
\widetilde{\rho}^{\,\sigma}\!\left(t,q\Bigl|{{\cal L}(q)\over
2q_1}\right)=\rho^{\sigma}\!\left(q\Bigl|{{\cal L}(q)\over
2q_1}\right)
\end{equation}
is time independent.

Finally we can write by using the identity
\beqs
\label{identity0}
{q_{1}^{4}+3q_{2}^{2}\over q_{1}}&=&6{\cal L}(p){{\cal L}(-q)
\over 2q_1}-2[3p_1{\cal L}(p)+2p_1^3]+\nonumber\\
&&2\left[3\!\left({{\cal L}(q)\over 2q_1}+q_1
+p_1 \right){\cal L}(p+q)+2(p_1+q_1)^3\right]
\eeqs
that
\beqs
&&\widetilde{v}(t,p)-v(p)=-{1\over 2}\!\int\!\!d^2q\,p_1
\sum_{\sigma =\pm}{\overline{\rho^{-\sigma}}
\!\left(q\bigl|{{\cal L}(q)\over 2q_1}\right)\over
q_1{\cal L}(p)+p_1{\cal L}(-q)-i\sigma 0q_1(q_1+p_1)}\times\\
&&\qquad\left[\exp\left(6it{\cal L}(p)\!\left({{\cal L}(p)\over2p_1}+
{{\cal L}(-q)\over2q_1}\right)\right)\widetilde{\rho}^{\,\sigma}
\!\left(t,p+q\Bigl|{{\cal L}(q)\over 2q_1}\right)-
\rho^{\sigma}
\!\left(p+q\Bigl|{{\cal L}(q)\over 2q_1}\right)\right].\nonumber
\eeqs
At the special values of $p$ and $q$ for which the denominator in the r.h.s.\
vanishes
\begin{equation}
{p+q\Bigl|}_{q_1{\cal L}(p)+p_1{\cal
L}(-q)=0}=\left\{\beta-\alpha,\beta^2-\alpha^2 \right\},
\end{equation}
with
\begin{equation}
\alpha={{\cal L}(q)\over2q_1}\,,\qquad\beta={{\cal L}(-p)\over-2p_1}
\end{equation}
and we obtain that, due to (\ref{stand:sd}) and (\ref{rhotilde}),
$\widetilde{\rho}$ and $\rho$ reduce to the same spectral data
\begin{equation}
\rho^\sigma(\beta-\alpha,\beta^2-\alpha^2|\alpha)\,.
\end{equation}
Therefore, at $q_1{\cal L}(p)+p_1{\cal L}(-q)=0$ the function in the square
brackets in the r.h.s.\ vanishes and $p_1$ can be extracted from the
integral. We conclude that
\begin{equation}
\lim_{p\to0}\widetilde{v}(t,p)=v(0).
\end{equation}
The continuity of $\widetilde{v}(t,p)$ at $p=0$ ensures us that the exponent
in (\ref{nonzerov}) is the dominant term in the time evolution of $v(t,p)$
and therefore we recover for $u(t,x,y)$ at large $x$ the behaviour computed
in the linearized case in (\ref{asymptotics}).

\subsubsection{$1/{\bf k}$--expansion in the inverse problem}

One can compute the coefficient of $1/{\bf k}$ in the asymptotic expansion
at large $\bf k$ of $\nu(t,p|{\bf k})$ also by using the Jost solution as
reconstructed from spectral data via the integral equation (\ref{IPcompl:t}).

Thanks to proposition~\ref{lemma2} in the appendix we have that
\beqs
&&\lim_{\hbox{\smallbf k}\to\infty}{\bf k}
\bigl[\nu(t,p|{\bf k})-\delta ^2(p)\bigr]=\nonumber\\
&&{1\over 4} \!\int\!\!{d^2q\,e^{it{q_{1}^{4}+3q_{2}^{2}\over q_{1}}}
\over q_1-i0t}\,\sum _{\sigma =\pm}
\overline{\rho^{-\sigma}} \Bigl(q\Bigl|{{\cal L}(q)\over 2q_1}\Bigr)\,
\nu^{\sigma}\Bigl(t,p+q\Bigl| {{\cal L}(q)\over 2q_1}\Bigr)\,.
\label{decomp:t}
\eeqs
As mentioned in the appendix the $i0t$ term is not relevant for $t\not=0$
and, therefore, we get from (\ref{v1t})
\begin{equation}
\label{dec1}
\lim_{\hbox{\smallbf k}\to\infty}{\bf k}\bigl[\nu(t,p|{\bf k})-
\delta ^2(p)\bigr]=-{v(t,p)\over 2p_1}\qquad \hbox{for}\,\,t\neq 0\,,
\end{equation}
The singular function $1/p_1$ does not need any regularization  at
$t\not=0$ since it is smoothed by the oscillating behaviour in time of
$v(t,p)$ indicated in
(\ref{nonzerov}). On the other side at $t=0$ (\ref {decomp:t}) is
discontinuous
\beqs
&&\lim_{t\to\epsilon 0}\lim_{\hbox{\smallbf k}\to\infty}{\bf k}
\bigl[\nu(t,p|{\bf k})-\delta ^2(p)\bigr]=\nonumber\\
&&{1\over 4}\sum_{\sigma =\pm}\!\int\!\!{d^2q\over q_1-i\epsilon 0}
\,\overline{\rho^{-\sigma}}\left(q\Bigl|{{\cal L}(q)\over 2q_1}\right)\,
\nu^{\sigma}\left(p+q\Bigl|{{\cal L}(q)\over 2q_1}\right)\,,
\qquad \epsilon =+,-\,.
\label{dec2}
\eeqs
Substituting  $(q_1-i\epsilon 0)^{-1}=(q_1\mp i{\cal L}(-q) 0)^{-1}+
2i\pi\epsilon\delta(q_1)\vartheta (\pm \epsilon q_2)$ we can use~(\ref {v4})
and
obtain
\beqs
&&\lim_{t\to\epsilon 0}\lim_{\hbox{\smallbf k}\to\infty}{\bf k}
\bigl[\nu(t,p|{\bf k})-\delta ^2(p)\bigr]=-{v(p)\over 2(p_1\pm ip_2
0)}+\nonumber\\
&&{i\pi\epsilon \over 2}\sum_{\sigma =\pm}\!\int\!\!dq_2
\lim_{q_1\to 0}
\,\overline{\rho^{-\sigma}}\left(q\Bigl|{{\cal L}(q)\over 2q_1}\right)\,
\nu^{\sigma}\left(p+q\Bigl|{{\cal L}(q)\over
2q_1}\right)\vartheta(\pm\epsilon q_2) \,,
\qquad \epsilon =+,-\,.
\label{dec3}
\eeqs
Due to~(\ref {sd0'}) and~(\ref {1'}) the last term is equal to
$i\pi\epsilon \vartheta (\mp\epsilon p_2)\delta(p_1)v(0,p_2)$ and,
consequently,
\begin{equation}\nonumber
\lim_{t\to\epsilon 0}\lim_{\hbox{\smallbf k}\to\infty}{\bf k}
\bigl[\nu(t,p|{\bf k})-\delta ^2(p)\bigr]=-{v(p)\over 2(p_1+i\epsilon0)}\,,
\end{equation}
Combining this result with the previous formula (\ref
{dec1}) we recover the
same coefficient (\ref {1t}) computed by using the integral equation for
$\nu(t,p|{\bf k})$ of the direct problem and, by recalling (\ref{decomp:t})
and (\ref{dec1}), we obtain that
\begin{equation}
\label{vtp/p1}
\frac{v(t,p)}{p_1+i0t}=-{1\over 2}
\!\int\!\!{d^2q\,e^{it{q_{1}^{4}+3q_{2}^{2}\over q_{1}}}
\over q_1-i0t}\,\sum _{\sigma =\pm}
\overline{\rho^{-\sigma}} \Bigl(q\Bigl|{{\cal L}(q)\over 2q_1}\Bigr)\,
\nu^{\sigma}\Bigl(t,p+q\Bigl| {{\cal L}(q)\over 2q_1}\Bigr)\,.
\end{equation}
\subsubsection{The recontructed time evolution of the Jost solution}

To derive the time evolution equation of the Jost solution we differentiate
with respect to $t$ equation~(\ref {IPcompl:t}). Proposition~\ref {lemma2}
in the appendix furnishes the derivative of the kernel of this equation
\beqs
{\partial\over \partial t}\,{e^{it{q_{1}^{4}+3q_{2}^{2}\over q
_{1}}} \over {\cal L}(-q)+2{\bf k}q_1}&=&
3i{\cal L}(-q){e^{it{q_{1}^{4}+3q_{2}^{2}\over q_{1}}} +
\over q_1-i0t}+\nonumber\\
&& 2i(2q_1^3+3(q_1-{\bf k}){\cal L}(-q))\,{e^{it{q_{1}^{4}+3q_{2}^{2}
\over q _{1}}} \over {\cal L}(-q)+2{\bf k}q_1} \,.
\label{der:t}
\eeqs
By using (\ref{vtp/p1}) and the identity
\begin{eqnarray}
\label{identity}
{q_{1}^{4}+3q_{2}^{2}\over q_{1}}&=&6{\cal L}(p)\!\left({\bf k}+{{\cal L}(-q)
\over 2q_1} \right)-2[3({\bf k}+p_1)\,{\cal L}(p)+2p_1^3]+\nonumber\\
&&2\left[3\!\left({{\cal L}(q)\over 2q_1}+q_1
+p_1 \right){\cal L}(p+q)+2(p_1+q_1)^3\right]
\end{eqnarray}
we obtain an integral equation of the type~(\ref {IPcompl:t}) for the quantity
$\partial \nu (t,p|{\bf k})/\partial t+2i\,\nu (t,p|{\bf k})$
$\bigl[3({\bf k}+p_1)\,{\cal L}(p)+2p_1^3\bigr]$ with the inhomogeneous term
equal
to $-3iv(t,p)\,{{\cal L}(p)\over p_1+i0t}$. Thanks to the assumption of
uniqueness of the  solution of this equation we get, finally, the
searched time evolution equation for the Jost solution
\begin{equation} \label{dernu}
{\partial \nu (t,p|{\bf k})\over \partial t}=
-2i\,\nu (t,p|{\bf k})\bigl[3({\bf k}+p_1)\,{\cal L}(p)+2p_1^3\bigr]-
3i\!\int\!\!d^2q\,v(t,q)\,{{\cal L}(q)\over q_1+i0t}\,\nu (t,p-q|{\bf k})\,.
\end{equation}
As expected the time derivative of $\nu(p|{\bf k})$ is discontinuous
at $t=0$
\begin{equation} \label{lim:dernu}
\lim_{t\to \pm 0}{\partial \nu (t,p|{\bf k})\over \partial t}=
-2i\,\nu (p|{\bf k})\bigl[3({\bf k}+p_1)\,{\cal L}(p)+2p_1^3\bigr]-
3i\!\int\!\!d^2q\,v(q)\,{{\cal L}(q)\over q_1\pm i0}\,\nu (p-q|{\bf k})\,.
\end{equation}
Note that at large $\bf k$ the leading term in the l.h.s.\ is of order
$1/{\bf
k}$ while in the r.h.s.\ it is of order ${\bf k}^0$ and, due to
(\ref{first}), it is
equal to
\begin{equation}
\label{0term}
3iv(p)\left({{\cal L}(p)\over p_1+i\sigma {\cal L}(p) 0}-
{{\cal L}(p)\over p_1\pm i0}\right)\,,\qquad \sigma =\sgn{\bf k}_{\Im}\,,
\end{equation}
This term, if the initial data $v(p)$ do not satisfy the constraint
$v(0,p_2)=0$, is zero only if $\sgn {\bf k}_{\Im}=\sgn(t{\cal L}(p))$. Once
more we recover the relevant fact that the two limits $t\to0$ and ${\bf
k}\to\infty$ can be exchanged only if condition (\ref {sign1}) is
satisfied.

\subsubsection{The reconstructed KPI equation and the
integrals of motion}

Recalling (\ref{defrho}), that we proved to be satisfied at all times, we
deduce from (\ref{dernu}) that
\beqs
&&{\partial \rho (t,p|{\bf k})\over \partial t}=
-6i\,{\cal L}(p){\bf k}\,\bigl[\rho (t,p|{\bf k})-v(t,p)\bigr]-\nonumber\\
&&3i{{\cal L}(p)^2\over p_1+i0t}\,v(t,p)-2ip_1\bigl[3{\cal L}(p)+
2p_1^2\bigr]\,\rho (t,p|{\bf k})-\nonumber\\
&&3i[{\cal L}(p)-2p_1{\bf k}]\!\int\!\!d^2q\,v(t,q)\,{{\cal L}(q)\over q_1+
i0t}\,\bigl[\nu (t,p-q|{\bf k})-\delta ^2(p-q)\bigr]\,.
\label{derrho}
\eeqs
The time evolution equation of $v(t,p)$ can be obtained from this equation
by computing the coefficient of the $1/{\bf k}$ term in its asymptotic
expansion at large $\bf k$.

We first note that from (\ref {1t}), by using (\ref {eqJ'}) that we proved
to be satisfied at all times, we get
\begin{equation}
\label{dec4}
\lim_{\hbox{\smallbf k}\to\infty}{\bf k}\bigl[\rho (t,p|{\bf k})-v(t,p)\bigr]
=-\!\int\!\!d^2q{v(t,p-q)\,v(t,q)\over 2(q_1+i0t)}\,.
\end{equation}
Then we multiply (\ref{derrho}) by $\bf k$ and compute the limit for ${\bf
k}\to\infty$ by inserting (\ref {1t}) and~(\ref {dec4}). We obtain
\beqs
&&{\partial v(t,p)\over \partial t}=
3i\,{\cal L}(p)\!\int\!\!d^2q{v(t,p-q)\,v(t,q)\over q_1+i0t}-\nonumber\\
&&i\left({3{\cal L}(p)^2\over p_1+i0t}+6p_1{\cal
L}(p)+4p_1^3\right)\,v(t,p)-\nonumber\\
&&3ip_1\!\int\!\!d^2q\,v(t,q)\,{{\cal L}(q)\over q_1+
i0t}\,{v(t,p-q)\over p_1-q_1+i0t} \,,
\eeqs
which can be easily transformed into the evolution form of the KPI equation
(\ref {kp2evol}) reported in the introduction.

Of special interest is the behaviour of
equation (\ref {derrho}) in a neighborhood of the point of discontinuity
$p=0$.  Precisely we consider the limit $p\to0$ along a path such that
$\lim_{p\to0}p_2^2/p_1=0$. Note that, because of (\ref{nonzerov}) and the
properties of
$\widetilde{v}(t,p)$ at $p=0$, only in this limit $v(t,0)=v(0)$.
Then all the
terms in the r.h.s.\ go
to zero and we
get that
\begin{equation} \label{derIM}
\lim_{p\to0\atop p_2^2/p_1\to 0}{\partial \rho (t,0|{\bf k})\over \partial
t}=0.
\end{equation}
{}From (\ref{IP:rho}), recalling (\ref{Sdt}), we get the dispersion relation
\begin{equation} \label{dispersion}
C({\bf k})\equiv\lim_{p\to0\atop p_2^2/p_1\to 0}\rho (t,0|{\bf
k})=v(0)+{1\over 2} \!\int\!\!
{d^2q\over {\cal L}(-q)+2{\bf k}q_1}\,\sum _{\sigma =\pm}
\overline{\rho^{-\sigma}}\Bigl(q\Bigl|{{\cal L}(q)\over 2q_1}\Bigr)\,
\rho^{\sigma}\Bigl(q\Bigl|{{\cal L}(q)\over 2q_1}\Bigr)\,.
\end{equation}
We conclude that $C({\bf k})$ generate an infinite number of conserved
quantities, that can be obtained by the recursion relations (\ref {recur})
computed at
$p=0$, when the limit is taken in the specified way. These conserved
quantities are expressed in terms of
the spectral data by means of equations (\ref{coeff1}) evaluated at $p=0$.



\section{Concluding Remarks}
\setcounter{equation}{0}

We conclude by attracting the attention of the reader to some specific
features of the formulation of the direct and inverse problems considered in
this article and in~\cite{kpshort}.
\begin{enumerate}
\item
The direct and inverse problems are formulated in terms of the Fourier
transformed Jost
solutions. It results that the corresponding variables $p_1$  and $p_2$ are
more proper for the
study of the properties of these solutions than the original ones $x$ and
$y$. For instance condition~(\ref {sign1}) which play a relevant role
in the theory cannot be reformulated
in terms of the $x$, $y$ variables.
\item
We use variables $p_1$ and $p_2$ and their combinations as arguments of the
spectral data instead of the standard variables $\alpha $ and $\beta $
(see~(\ref {subst}) and (\ref{stand:sd})). These variables are suggested by
the form of the
direct problem. Properties of the spectral data are more naturally
formulated in
their terms.
\item
The main tool of the inverse problem is a set of two integral equations
for the boundary values of the Jost solutions on the real axis of the
complex plane of the spectral parameter $\bf k$. This set  depends linearly
and not quadratically, as in the standard formulation, on the spectral data.
In addition  the spectral data
themselves are given in terms of the Jost solutions by means of~(\ref
{def:sd}). This closed formulation linear in the spectral data enables us,
for instance, to
study easier the properties of the
Jost solutions under time evolution and to deduce easier the asymptotic
expansion at large $\bf k$ and dispersion relation~(\ref
{dispersion}).
\end{enumerate}
Functional character of spectral data and their different properties
according to the choice of spectral variables require some special remarks.

If we choose the standard set of spectral data~(\ref
{stand:sd}) we get spectral data which have bad behaviour for $\alpha $ and
$\beta $
going to infinity. More precisely, due to~(\ref {sd0}), (\ref {sd0'}), (\ref
{rsd0'}) and (\ref {rsd0''}),
the diagonal values $r^{\pm}(\beta,\beta)$ and values `close' to the
diagonal do not decrease
at infinity (see also~\cite{kpshort}). It is necessary to emphasize that
this happens for initial data of KPI as small in norm and smooth as we like.
The
only possibility to remove this constant behavior is to impose the
constraint $v(0,p_{2})=0$ (i.e.~(\ref {t0constraint})) on initial data. In
this case we get a $1/\beta $ behavior,
to improve which we need a second constraint and so on. Due to~(\ref {sd0}) it
is clear that in order to get spectral data $r^{\pm}(\beta,\beta)$ decreasing
for $\beta\to\infty $ quicker then any negative power of $\beta $ it is
necessary to impose an  infinite set of constraints on initial data. Let us
mention that this situation is specific of the $2+1$ dimensional case and
has no
analogue in $1+1$.

In~\cite{kpshort} it was shown that the study of the properties of the Jost
solutions by means of the inverse problem in terms of spectral data
$r^{\pm}(\alpha,\beta)$ needs a special care due to this asymptotic
behavior. In order to compute the asymptotic $1/{\bf k}$--expansion or the
time derivative
of the Jost solutions it is necessary to use a special subtraction
procedure to avoid divergent expressions.

The use of the spectral data $\rho ^{\pm}\left(p,{{\cal L}(p)\over 2p_{1}}
\right)$ in this article enables us to remove this complications.
This functions rapidly decrease for $p\to\infty$ in all directions (roughly
speaking as initial data $v(p)$) and are smooth at $p_{1}\not=0$ (see~(\ref
{sd})). Of course difficulties cannot disappear thanks to a convenient
change of variables and in fact also the spectral data $\rho
^{\pm}\left(p,{{\cal L}(p)\over 2p_{1}}\right)$ do not belong to the
Schwartz space
since they are discontinuous at $p=0$. This difficulty is, however, more
naturally manageable and direct and inverse problems can be formulated
practically in the standard way. Special attention must be paid to the
discontinuity of the time derivatives at $t=0$. This problem is controlled by
the lemma and propositions given in the appendix.

As shown in section 3 the properties of all the quantities under
consideration are essentially different for $t=0$ and $t\not=0$. This
situation is specific of the multi--dimensional integrable evolution
equations, which are non local already in their
linearized version. In fact the nonlocality implies that the dispersion
relation of spectral data
is singular as indicated for instance in~(\ref {Sdt}). Then for $t\not=0$
the rapid oscillations of the exponent in~(\ref
{Sdt}) near
$p_1=0$ can control a singularity at $p_1=0$ while for $t=0$ a specific
regularization is needed. This is hidden if we use
the standard spectral data $r^{\pm}(\alpha,\beta)$. In fact the
exponent in~(\ref {derSD2'3}) changes the functional class of the
initial
data. The diagonal values
$r^{\pm}(t,\beta,\beta)$ are time independent and thus have the same bad
behaviour at large $\beta$ as the spectral data at time $t=0$, but
asymptotic expansions `close' to
the diagonal of the type~(\ref {rsd0''}) show again rapid oscillations
for $t$ and $p_2$ different from zero.


\paragraph{Acknowledgments}
A.~P. thanks his colleagues at Dipartimento di Fi\-si\-ca
dell'Uni\-ver\-sit\`a di Lecce for kind hospitality.

\appendix

\section{Appendix}\label{lemmas}
\setcounter{equation}{0}

\begin{lemma}\label{lemma1}
The function $\exp(i\tau / p_1)$ defines a distribution in the Schwartz
space of the variable $p_1$ depending continuously on the
parameter $\tau $.
\begin{enumerate}
\item[{\sl i)}] This distribution is continuously differentiable
in $\tau $ for any $\tau \neq 0$. Precisely, we have
\begin{equation}
\label{deriv1}
\partial_{\tau }\exp\left({i\tau \over p_1}\right)=
{i\over p_1} \,\exp\left({i\tau \over p_1}\right),\qquad \tau \neq 0,
\end{equation}
where the r.h.s.\ is a well defined distribution in the same space.
\item[{\sl ii)}] At $\tau =0$ there exist right/left limits
\begin{equation}
\lim_{\tau \to \pm 0}{1\over p_{1}}\,
\exp\left({i\tau \over p_1}\right)= {1\over p_1\mp i0}.\nonumber
\end{equation}
\end{enumerate}
Thus $\partial_{\tau }\exp\left({i\tau \over p_1}\right)$ can be considered
a distribution also in the variable $\tau$ and we can write for arbitrary
$\tau $
\begin{equation}
\label{derivexp}
\partial_{\tau }\exp\left({i\tau \over p_1}\right)=
{i\over p_1-i0\tau } \,\exp\left({i\tau \over p_1}\right).
\end{equation}
\end{lemma}
{\sl Proof}

{\sl i)} Let us consider
\beqsn
f(\tau )=\!\int\!\! dp_1\,e^{i\tau/p_1}
\varphi (p_1)
\eeqsn
where $\varphi (p_1)\in {\cal S}$. Subtracting and adding terms we can write
\beqsn
f(\tau )=\!\int\!\!dp_1\,e^{i\tau/p_1}
\bigl[ \varphi (p_1)-\vartheta (1-| p_1| ) \varphi (0) \bigr]
+\varphi (0)\!\int^{1}_{-1}\!\!dp_1 \,e^{i\tau/p_1}.
\eeqsn
Changing $p_1$ to $1/p_1$ in the second term we have
\beqsn
f(\tau)=\!\int\!\!dp_1\,e^{i\tau/p_1} \bigl[
\varphi (p_1)-\vartheta (1-| p_1| ) \varphi (0) \bigr] +
\varphi (0)\!\int_{|p_1|\geq1}\!\!{dp_1\over p_1^2}  \,e^{i \tau
p_1 }.
\eeqsn
Now it is obvious that both terms are differentiable in $\tau $
for $\tau \neq 0$ and
\beqsn
\partial_{\tau }f(\tau )=i\!\int\!\!{dp_1 \over p_1}\,
e^{i\tau/p_1} \bigl[ \varphi (p_1)-\vartheta (1-|
p_1| ) \varphi (0) \bigr] +
i\varphi (0)\!\int_{|p_1|\geq1}\!\!{dp_1\over p_1}\,
e^{i\tau p_1 } .
\eeqsn
Substituting again $p_1$ for $1/p_1$ in the second term we obtain
that it cancels out with the second term in brackets.  This
proves {\sl i)}. Let us remark that the last formula shows that for
$\tau \neq 0$ it is  not necessary to regularize the factor $1/p_1$ in
front of the exponent in (\ref {deriv1}). As well one can consider this
factor indifferently as principal value or as $(p_1\pm i0)^{-1}$.

{\sl ii)} Again subtracting and adding terms we can write
\beqsn
\!\int\!\! {dp_1\over p_1} \,e^{i\tau/p_1}\,
\varphi (p_1) &=&
\int  {dp_1\over p_1} e^{i\tau/p_1} \left[
\varphi (p_1)-\vartheta (1-| p_1| ) \varphi (0) \right] +\\
&&\sgn \tau \,\varphi (0)\int_{|p_1|\geq|\tau |} {dp_1\over p_1} \,e^{i
p_1} ,
\eeqsn
where now in the second term $p_1$ was substituted for $\tau p_1$.
To compute the limit for $\tau \to 0$ notice that
\beqsn
\lim_{\tau \rightarrow  0}\int_{| p_1|\geq| \tau | }
{dp_1\over p_{1}}\,e^{ip_1} =
i\pi.
\eeqsn
Thus
\beqsn
\lim_{\tau \rightarrow  0}\int {dp_1\over p_1}\,
e^{i\tau/p_1} \varphi (p_1)& =&
\int  {dp_1\over p_1} \left[ \varphi (p_1)-\vartheta
(1-| p_1| ) \varphi (0) \right] \pm i\pi
\varphi (0)\equiv\\
&&\int  {dp_1\over p_1} \, \varphi (p_1)\pm i\pi\varphi (0),
\eeqsn
where to get the second equality we chose $p_1^{-1}$ in the
integral as the principal value. Then (\ref {derivexp}) follows due
to (\ref {deriv1}) and the remark made after the proof of
{\sl i)}.

Some useful propositions follow  from this lemma.
\begin{prop}\label{lemma2}
The function ${\displaystyle e^{-3it{p^{2}_{2}\over p_{1}} }
\over p_{2}-2{\bf k}p_{1}}$
defines a distribution in the
Schwartz space of the variables $p_1$, $p_2$
 depending continuously on the parameter $t$.
\begin{enumerate}
\item[{\sl i)}] This distribution is continuously differentiable
in $t$ for any $t\neq 0$ and has well defined right/left limits at $t=0$
according to the following formula
\begin{equation}
{\partial \over \partial t} {e^{-3it{p^{2}_{2}\over p_{1}} } \over
p_{2}-2{\bf k}p_{1}} =-{6i{\bf k}p_{2}\over p_{2}-2{\bf k}p_{1}} e^{-3it
{p^{2}_{2}\over p_{1}} }- {3 ip_{2}\over p_{1}+i0t} e^{-3i t {p^{2}_{2}
\over p_{1}} }
\end{equation}
\item[{\sl ii)}]
\begin{equation}
\lim_{{\bf k}\rightarrow \infty }{{\bf k} \over p_{2}-2{\bf k}p_{1}}
e^{-3i t {p^{2}_{2}\over p_{1}} } =- {1\over 2} {e^{-3i t {p^{2}_{2}
\over p_{1}} }\over p_{1}+i0t}
\end{equation}
\end{enumerate}
\end{prop}

Note that, due to (\ref{expan}),
\begin{equation}
\lim_{\hbox{\smallbf k}\rightarrow \infty }{{\bf
k} \over p_{2}-2{\bf k}p _{1}}  = -{1\over
2(p_{1}+i0\sigma p_{2})},\qquad \sigma =\sgn {\bf k}_{ \Im},
\end{equation}
and therefore we get from the previous proposition:
\begin{prop}\label{lemma3}
The limits in the following formula
\begin{equation}
\lim_{t\to\pm0}\lim_{\hbox{\smallbf k}\rightarrow \infty }{{\bf
k} \over p_{2}-2{\bf k}p _{1}} e^{-3i t {p^{2}_{2}\over p_{1}} }
\end{equation}
can be exchanged if and only if
\begin{equation}
\sgn {\bf k}_{ \Im}=\sgn{tp_2}.
\end{equation}
\end{prop}
{}From the point $ii)$ of lemma \ref{lemma1} we get that
\begin{prop}
For any distribution defined by
\begin{equation}
U(t,x,y) = \!\int\!\! d^{2}p\, e^{-ip_{1}x+ip_{2}y}
\exp\left(-it\,{p_{1}^{4}+3p_{2}^{2}\over p_{1}}\right)\,V(p)
\end{equation}
with $V(p)$ an arbitrary function belonging to the Schwartz space
\beqs
&&\lim_{t\to\pm0}\int_{\infty}^{x}\!\!dx'U(t,x',y)=
\!\!\int_{\mp\infty}^{x}\!\!dx' U(0,x',y)\\
&&\lim_{t\to\pm0}\int_{-\infty}^{x}\!\!dx'U(t,x',y)=
\!\!\int_{\mp\infty}^{x}\!\!dx' U(0,x',y).
\eeqs
Therefore the  operator inverse of $\partial_x$ defined by
$\int_{-t\infty}^{x}\!\!dx'$  commute with the limit $t\to0$
\begin{equation}
\lim_{t\to\pm0}\int_{-t\infty}^{x}\!\!dx'U(t,x',y)=
\!\!\int_{\mp\infty}^{x}\!\!dx' U(0,x',y)
\end{equation}
while the antisymmetrical inverse operator satisfies
\begin{equation}
\lim_{t\to\pm0}{1\over2}\left(\int_{-\infty}^{x}\!\!{+}\!\! \int_{+\infty}^{x}
\right)\!dx'U(t,x',y)=
\!\!\!\int_{\mp\infty}^{x}\!\!dx' U(0,x',y).
\end{equation}
\end{prop}


\end{document}